\documentclass[
 reprint,
 superscriptaddress,
preprintnumbers,
 pra,
showkeys
]{revtex4-2}

\usepackage{float}
\usepackage{multibib}
\usepackage{hyperref}
\usepackage{xcolor}
\hypersetup{
    colorlinks,
    linkcolor={blue!50!blue},
    citecolor={blue!50!blue},
    urlcolor={blue!80!black}
}
\usepackage{import}
\usepackage{placeins}
\usepackage{xcolor}

\usepackage{lmodern}
\usepackage{amssymb}

\usepackage{graphicx}
\usepackage{dcolumn}
\usepackage{bm}
\usepackage{physics}  

\begin{document}

\title{Polarization dynamics of solid-state quantum emitters}  

\author{Anand Kumar}
\thanks{These authors contributed equally.}
\affiliation{Department of Computer Engineering, School of Computation, Information and Technology, Technical University of Munich, 80333 Munich, Germany}
\affiliation{Abbe Center of Photonics, Institute of Applied Physics, Friedrich Schiller University Jena, 07745 Jena, Germany}
\author{\c{C}a\u{g}lar Samaner}
\thanks{These authors contributed equally.}
\affiliation{Department of Physics, \.{I}zmir Institute of Technology, 35430 \.{I}zmir, Turkey}
\author{Chanaprom Cholsuk}
\affiliation{Department of Computer Engineering, School of Computation, Information and Technology, Technical University of Munich, 80333 Munich, Germany}
\affiliation{Abbe Center of Photonics, Institute of Applied Physics, Friedrich Schiller University Jena, 07745 Jena, Germany}
\author{Tjorben Matthes}
\affiliation{Department of Computer Engineering, School of Computation, Information and Technology, Technical University of Munich, 80333 Munich, Germany}
\affiliation{Abbe Center of Photonics, Institute of Applied Physics, Friedrich Schiller University Jena, 07745 Jena, Germany}
\author{Serkan Pa\c{c}al}
\affiliation{Department of Physics, \.{I}zmir Institute of Technology, 35430 \.{I}zmir, Turkey}
\author{Ya\u{g}{\i}z Oyun}
\affiliation{Department of Photonics, \.{I}zmir Institute of Technology, 35430 \.{I}zmir, Turkey}
\author{Ashkan Zand}
\affiliation{Department of Computer Engineering, School of Computation, Information and Technology, Technical University of Munich, 80333 Munich, Germany}
\affiliation{Abbe Center of Photonics, Institute of Applied Physics, Friedrich Schiller University Jena, 07745 Jena, Germany}
\author{Robert J. Chapman}
\affiliation{Optical Nanomaterial Group, Institute for Quantum Electronics, Department of Physics, ETH Zurich, 8093 Zurich, Switzerland}
\author{Gr\'{e}goire Saerens}
\affiliation{Optical Nanomaterial Group, Institute for Quantum Electronics, Department of Physics, ETH Zurich, 8093 Zurich, Switzerland}
\author{Rachel Grange}
\affiliation{Optical Nanomaterial Group, Institute for Quantum Electronics, Department of Physics, ETH Zurich, 8093 Zurich, Switzerland}
\author{Sujin Suwanna}
\affiliation{Optical and Quantum Physics Laboratory, Department of Physics, Faculty of Science, Mahidol University, 10400 Bangkok, Thailand}
\author{Serkan Ate\c{s}}
\thanks{\href{mailto:serkanates@iyte.edu.tr}{serkanates@iyte.edu.tr} and \href{mailto:tobias.vogl@tum.de}{tobias.vogl@tum.de}.}
\affiliation{Department of Physics, \.{I}zmir Institute of Technology, 35430 \.{I}zmir, Turkey}
\author{Tobias Vogl}
\thanks{\href{mailto:serkanates@iyte.edu.tr}{serkanates@iyte.edu.tr} and \href{mailto:tobias.vogl@tum.de}{tobias.vogl@tum.de}.}
\affiliation{Department of Computer Engineering, School of Computation, Information and Technology, Technical University of Munich, 80333 Munich, Germany}
\affiliation{Abbe Center of Photonics, Institute of Applied Physics, Friedrich Schiller University Jena, 07745 Jena, Germany}

\date{\today}

\begin{abstract}
  Quantum emitters in solid-state crystals have recently attracted a lot of attention due to their simple applicability in optical quantum technologies. The polarization of single photons generated by quantum emitters is one of the key parameters that play a crucial role in the applications, such as quantum computation that uses the indistinguishability of photons. However, the degree of single photon polarization is typically quantified using time-averaged photoluminescence intensity of single emitters, which provides limited information about the dipole properties in solids. In this work, we use single defects in hexagonal boron nitride and nanodiamond as efficient room-temperature single photon sources to reveal the origin and the temporal evolution of dipole orientation in solid-state quantum emitters. The angle of excitation and emission dipoles relative to the crystal axes are determined experimentally and then calculated using density functional theory, which results in characteristic angles for every specific defect that can be used as an efficient tool for defect identification and understanding their atomic structure. Moreover, the temporal polarization dynamics reveal a strongly modified linear polarization visibility that depends on the excited state decay time of individual excitation. This effect can be traced back potentially to the excitation of excess charges in the local crystal environment. Understanding such hidden time-dependent mechanisms can further be used to improve the performance of polarization-sensitive experiments, in particular that of quantum communication with single photon emitters.
\end{abstract}


\maketitle

\section{Introduction}
Fluorescent defects in solid-state crystals have become one of the most promising sources of single photons \cite{10.1038/nphoton.2016.186} for near-future quantum information processing and integrated quantum photonics \cite{Atature2018}. The key roles of single photon sources for optical quantum computing \cite{Quantumcomp}, quantum key distribution (QKD) \cite{Lo2014}, nanoscale quantum sensors \cite{Degen2017}, and fundamental quantum optics experiments \cite{PhysRevResearch.3.013296} have fueled the research on quantum emitter (QE) systems. Particularly important for these photon sources is the stability of their intrinsic properties. An unsteady polarization for example limits the coherence (for interferometry), indistinguishability (for optical quantum computing), or entropy (for QKD) of a photon source. There have been many material platforms identified that can host stable single photon emitters at room temperature, including hexagonal boron nitride (hBN) \cite{nnano.2015.242}, diamond \cite{Bradac2019}, silicon nitride \cite{doi:10.1126/sciadv.abj0627}, and zinc oxide \cite{doi:10.1021/nl204010e} to name only a few. All these systems have in common that a point-like defect induces additional energy levels into the wide band gap of the host materials.\\
\indent Due to the relatively recent discovery of fluorescent defects in hBN \cite{nnano.2015.242}, their atomic structures are not well understood compared to established emitter systems \cite{C9NR04269E}, such as the nitrogen-vacancy (NV) centers in diamond \cite{DOHERTY20131}. Defects in hBN can be created artificially via localized electron and ion implantation \cite{10.1021/acsami.6b09875}, oxygen plasma treatment \cite{10.1021/acsphotonics.8b00127}, chemical etching \cite{10.1021/acs.nanolett.6b03268}, gamma-rays \cite{10.1038/s41467-019-09219-5}, and activated through strain \cite{10.1364/OPTICA.5.001128}. It is possible to control the defect formation such that emitter arrays can be fabricated using nanoindentation with an atomic force microscope \cite{doi:10.1021/acs.nanolett.1c02640} and localized electron irradiation \cite{naturecomm,Gale_2022,akumar}.\\

\begin{figure*}[t]
    \makebox[\textwidth]{\includegraphics[width=.9\paperwidth]{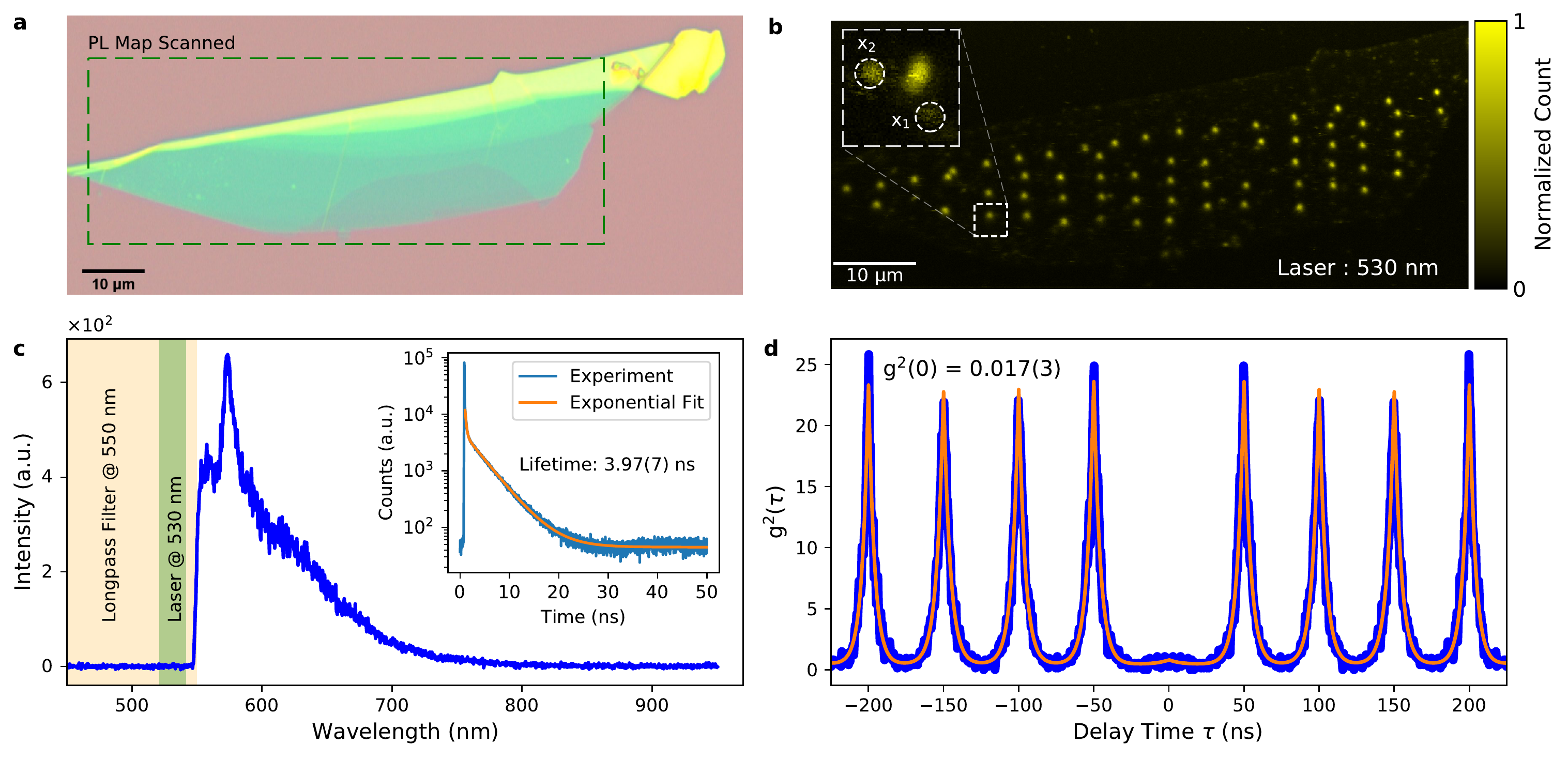}}
    \caption{\textbf{a} Optical microscope image of the exfoliated hBN flake on a Si/SiO$_2$ substrate. \textbf{b} PL maps of the irradiated array, excited with a 530 nm pulsed laser at a repetition rate of 20 MHz. The inset image is a zoomed-in PL map of one of the irradiated spots revealing multiple single emitter spots. \textbf{c} Typical spectrum of a single emitter with a peak of the PL at 575 nm and detected with a long-pass filter at 550 nm that cuts the emission partially. The inset figure shows the typical lifetime decay curve revealing a lifetime of 3.96(7) ns. \textbf{d} The second-order correlation function under pulsed excitation at the position marked `x$_1$' in \textbf{b}, with $g^{(2)}(0)= 0.017(3)$ and `x$_2$' $g^{(2)}(0)=0.042(2)$. The $g^{(2)}(0)$ values are extracted from the fitted curve.}
    \label{fig-1}
\end{figure*}

\indent Recent investigations have identified the negatively charged boron vacancy as the near-infrared emitter through optically detected magnetic resonance measurements \cite{Gottscholl2020}. Other experiments have linked carbon impurities to the visible emitters in the blue \cite{Gale_2022} and green-red \cite{Mendelson2021} regions of the spectrum. These works have used magnetic or spectral properties for emitter identification. Another option could be using the dipole polarization dynamics, which is also characteristic of every specific defect. This is, however, only meaningful when a large number of identical emitters are investigated, and sufficient statistics are collected. The fabrication of such identical emitter arrays has been achieved recently \cite{naturecomm,Gale_2022,akumar}. The systematic study of the emission dipole angles remained inconclusive, as either only a few emitters were studied \cite{naturecomm, Jungwirth} or the dipoles were randomly distributed \cite{PhysRevApplied.18.064021}. The latter could indicate a surface complex that does not form a chemical bond with the hBN lattice and therefore can be oriented randomly. Vacancy-related defects in wrinkled hBN have shown a strong correlation of the polarization axes with the wrinkle direction in the crystal \cite{yim_et_al}. The various experimental and theoretical models have advanced the insight into the emitters, yet remained elusive to identify the defect \cite{naturecomm, PhysRevApplied.18.064021, Mendelson2020, Ivady2020, tawfik_ali_fronzi_kianinia_tran_stampfl_aharonovich_toth_ford_2017}.\\
\indent When such defect-based emitters are used in quantum communication scenarios and information is encoded in the polarization, recent studies demonstrated a performance improvement of the quantum communication protocols by temporal filtering and post-selection \cite{Ates2013, Caglar}. This was always a well-known effect due to detector dark counts that can be suppressed this way. As the emission process of a fluorescent defect is usually complex, it is important to understand the emission dynamics, which can potentially enhance the performance in quantum technology applications even further. There have been some insights into the dynamics of the optical transitions from multiple electronic excited states in hBN \cite{Patel2022}, but the relation to the transition dipole moments is still missing.\\  
\indent In this work, we study the polarization dynamics of a large array of identical `yellow' quantum emitters. We investigate the correlation of excitation and emission polarization with the host crystal axes. Our experiments are supported by density functional theory (DFT) calculations, which can model these dipole characteristics that are a characteristic fingerprint of any specific defect. Furthermore, we also time-resolve our polarization measurements to gain insight into the emission mechanism. This is generalized to other samples containing quantum emitters, including hBN nanoflakes and NV centers in diamonds. We, therefore, provide important insights into the polarization dynamics of general solid-state quantum emitters. This includes the oddity of misaligned excitation and emission dipoles, non-unity polarization visibility, as well as the atomic structure of the yellow hBN emitter.

\section{Results and discussion}
A thin hBN flake is mechanically exfoliated from bulk crystal to a silicon substrate with a 298 nm thick thermal oxide layer (see Methods).  Optically active emitters are induced using localized electron beam irradiation with a standard scanning electron microscope (SEM) at a chosen spot in the flake \cite{akumar}. The emitters are created over the entire flake, independently of the local flake thickness (see Supplementary Section S1.1). Fig.\ \ref{fig-1}a shows an optical microscope image of the flake and Fig.\ \ref{fig-1}b the resulting photoluminescence (PL) map under pulsed laser excitation with a 530 nm laser (see Methods). The PL map reveals diffraction-limited emission spots that originate from quantum emitters formed during the irradiation and defect formation process. The inset in Fig.\ \ref{fig-1}b shows a zoomed-in PL map of one of the irradiated spots where more than one emitter is present. This is due to the probabilistic nature of the irradiation process. A typical emission spectrum is shown in Fig.\ \ref{fig-1}c with a peak at 575 nm (see Supplementary Section S1.2 for the statistics on the spectra within the array). The typical lifetime of the emitter is around 4 ns (see inset). Due to the used long-pass filter that suppresses the excitation laser, we do not have full access to the spectrum. We still expect that there is not much emission below 550 nm as the excitation with a 470 nm laser in a separate measurement was very inefficient (see supplementary Section S1.3), implying no available phonon modes in the blue. The polarization-resolved second-order correlation measurement in Fig.\ \ref{fig-1}e proves single photon emission with $g{^2}(0)= 0.0171(3)$ for the spot labeled `x$_1$' in Fig.\ \ref{fig-1}b (and $0.0410(6)$ for `x$_2$') without any background correction (see Supplementary Section S1.4). The polarization-resolved measurements allow us to selectively excite the emitter efficiently by matching the laser polarization and at the same time suppress any uncorrelated noise sources nearby. This in turn leads to a better single photon purity compared to measurements with a fixed (but random relative) polarization as in previous experiments \cite{akumar}. All emitters in the array have near-identical photophysical properties (see Supplementary Section S1 and the following sections), which allow us to study the polarization dynamics.

\begin{figure*}[t]
    \makebox[\textwidth]{\includegraphics[width=.9\paperwidth]{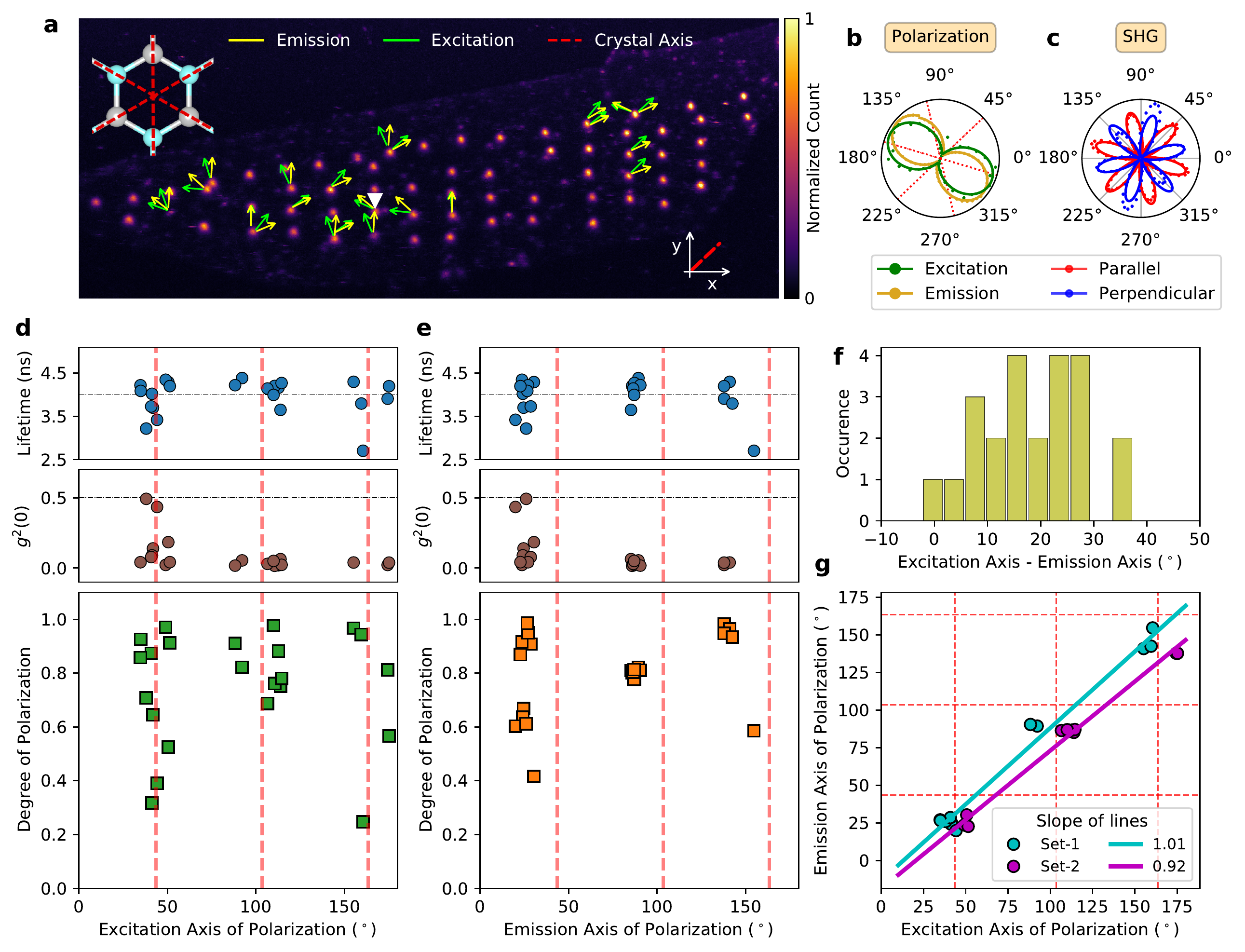}}
    \caption{\textbf{a} PL map of the entire flake created using a pulsed excitation laser at 530 nm. The emission and excitation axes of measured emitters are presented with arrows at the measured angle relative to the (random) x-axis as marked in the map. One of the main crystal axes has an angle of 43.52$^\circ$ $\pm$ 0.39$^\circ$ with respect to the x-axis. \textbf{b} A typical polar plot of emission and excitation axis at the spot marked with a white `$\blacktriangledown$' symbol. The degree of polarization is extracted from a cosine-squared fit with 98.01\% (emission) and 96.67\% (excitation). Here `red' and `blue' grid lines present the crystal axis in order to correlate the emission and excitation axis with respect to the crystal axis. \textbf{c} The polarization-resolved SHG measurement reveals the crystallographic axes, as evident by the six-fold symmetry. These axes are also marked in all sub-plots. The scatter plot of the measured excitation and emission axes against the degree of polarization of the emitters are shown in \textbf{d} and \textbf{e}, respectively. All the emitters presented in the scatter plot have a clear g$^2(\tau = 0)$ dip and an average lifetime of around 4 ns as indicated in the plot. \textbf{f} The misalignment between excitation and emission axis of polarization with a mean value of 18.9(100)$^\circ$. \textbf{g} Emission versus excitation axes showing a linear behavior. The plot is showing a clear splitting into two groups identified as `Set-1' and `Set-2' with a slope of nearly one.}
    \label{fig-2}
\end{figure*}

\subsection{Correlation of emitter polarization with crystal axis}
To study the excitation dipole axes of our emitters, we first polarized our laser circularly using a quarter-wave plate. The actual excitation laser polarization is then set using a linear polarizer with a high extinction ratio. The initial circular polarization ensures equal excitation power independent of the current polarization. This combination yields a more accurate excitation polarization compared to a simple half-wave plate (see Supplementary Section 2.1). The laser power is monitored using power-meter with a variation below 5\%. The polarizer is rotated from 0$^\circ$ to 360$^\circ$ in steps of 10$^\circ$ or 15$^\circ$ using a motorized mount. A fitting routine of the data allows us to extract the polarization directions with much higher accuracy than the rotation step size (see Supplementary Section 2.2). We have noticed a small beam shift in the PL map during the rotation, likely caused by the optical component being not plane-parallel. Instead of simply recording the PL count rate using the single photon avalanche diodes (SPADs), we recorded local PL maps and integrated the intensity of the diffraction-limited spots (see Supplementary Section S2.3). This way we compensate for the slight beam shift.  In the detection path, we have another motorized polarizer to measure the emission dipole axes. Note that this polarizer was only present during the emission dipole measurements and not for the excitation measurements.\\
\indent For measuring the emission dipoles, the laser polarization is set to have a maximal overlap with the excitation dipole. Once the excitation polarizer is optimized, we record the time trace (PL signal over time) while rotating the polarizer with a dwell time of 5 s in the detection path. Afterward, we extracted the integrated PL intensity (see Supplementary Section S1.4). In some cases, we observed multiple polarization axes which could be due to the presence of multiple emitters within a diffraction-limited spot (see Supplementary section S1.5). However, we have omitted such cases from our analysis and only consider emitters with a clear $g^{(2)}(0)$ dip and unique polarization axes.\\
\indent Note that we always specify the polarization axis, not the dipole axis (which is rotated by 90$^\circ$ from the polarization axis). The PL map with green/yellow arrows respectively marking the excitation/emission polarization axis is shown in Fig.\ \ref{fig-2}a as extracted from the fitting routine. Fig.\ \ref{fig-2}b shows a typical polar plot of the measured and fitted data from an emitter marked with a triangle in  Fig.\ \ref{fig-2}a. We also extracted the error bars from our fitting which turn out to be smaller than the symbol size and thus omitted in Fig.\ \ref{fig-2} (more details can be found in Supplementary Section S3.1). We also observed that the orientation of the excitation/emission axis is independent of the flake thickness (see Supplementary Section S3.2). This distribution of polarization axes is expected, due to the specific order of layer stacking in (crystalline) hBN \cite{Constantinescu2013}. The layer orientation or individual local flake thickness has therefore no influence on the polarization axes of the defect centers. \\
\indent The question arises about how the dipole aligns within the crystal lattice. This can be easily probed with polarization-resolved second-harmonic generation (SHG) \cite{doi:10.1021/nl401561r}, which reveals the sixfold axis symmetry of the hBN lattice (see Fig.\ \ref{fig-2}c). The quadratic pump power dependency verifies the second-order process of SHG (see Supplementary Section S4). The crystallographic axes are also indicated in Fig.\ \ref{fig-2}a and b by the dashed red lines.\\
\indent We have been able to record both excitation and emission dipoles for 23 emitters. The scatter plots of the excitation and emission angle (modulo 180$^\circ$) distribution are shown in Fig.\ \ref{fig-2}d and e. For every emitter, we have also measured the g$^{(2)}$(0)-value and the excited state lifetime to verify that we have single emitters, which relax through the same decay channel (see the top part of Fig.\ \ref{fig-2}d and \ref{fig-2}e, and also Supplementary Sections S1.5 and S1.6 for the raw data). We also extract the degree of linear polarization from our fits and display this in the histogram. It is worth noting that with our measurements, we have only projected onto the equatorial plane of the Poincar\'{e} sphere. A full quantum state tomography would require projecting onto the circular components as well. For simplicity, we will refer to the degree of linear polarization simply as the degree of polarization. Many emitters have a high polarization visibility above 80\%. It is clear that for the yellow emitters, the excitation and emission axes bunch around certain angles in relation to the crystal axis with an uncertainty range of 8$^\circ$ for excitation and 4$^\circ$ for emission. The misalignment between excitation and emission is on average 18.9(100)$^\circ$ as shown in Fig.\ \ref{fig-2}f. This large uncertainty makes it difficult to assign a specific defect complex (which we, therefore, do not attempt). In general, this could be due to multiple involved transitions or local modifications in the crystal environment and needs further investigation.\\
\indent We also observe a splitting between the excitation and emission axes of polarization, and this becomes even clearer when the emission axis is plotted versus the excitation axis in Fig.\ \ref{fig-2}g, where actually six groups can be distinguished. If the excitation/emission polarization co-aligns with a crystal axis or is exactly in the middle of the crystal axis, a threefold symmetry results (i.e., three groups). If there is an angle between the polarization and crystal axis, these three groups split into six symmetrically around the crystal axis. In our case, however, the centers of the groups are not separated by 60$^\circ$ (for emission), and the mean distances from the crystal axis range from roughly 3 to 10$^\circ$ (see Supplementary Section S5). Moreover, the splitting in the emission polarization is less prominent and not symmetric around a crystal axis (unlike that in the excitation polarization). This symmetry breaking could be due to localized strain in the crystal lattice induced during the localized electron irradiation process or other local modifications of the crystal environment. It is important to note that we have observed some anisotropy in our SHG measurement in Fig.\ \ref{fig-2}c which is related to residual strain in the crystal lattice \cite{Mennel2018}. However, this is the global strain that is typically induced during the exfoliation process. The local strain, in particular around the irradiated spots could be considerably higher and is not resolvable with our SHG setup.  Such strain can also lead to the change in polarization axis \cite{Mendelson2020}, and to further investigate it, we model this qualitatively using DFT.

\subsection{Temporal polarization dynamics}

\begin{figure*}[t]
    \makebox[\textwidth]{\includegraphics[width=1.05\textwidth]{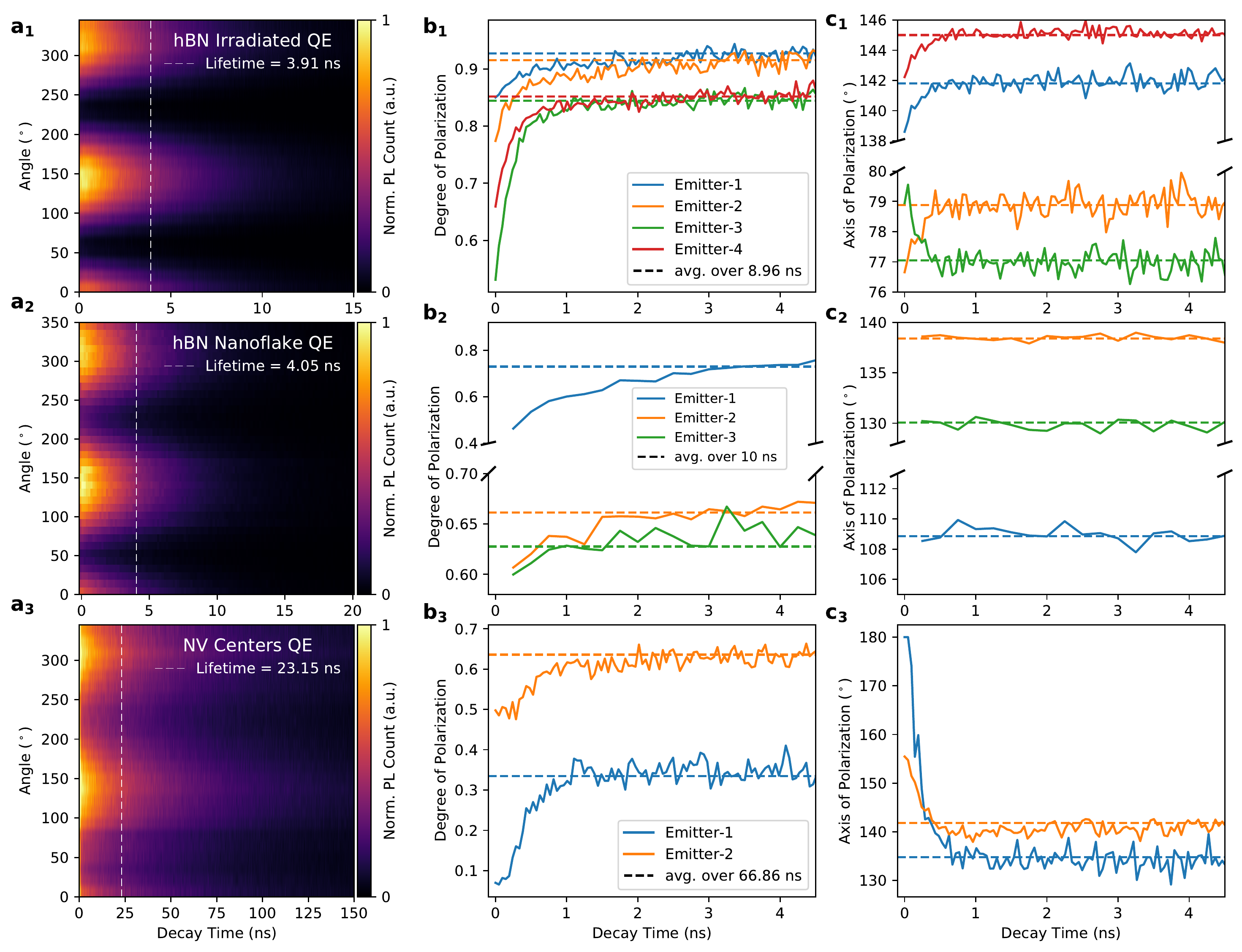}}
    \caption{\textbf{a$_\mathbf{i}$} The PL intensity with respect to time after the excitation laser pulse as function of polarizer rotation angle measured for \textbf{a$_1$} hBN irradiated QEs, \textbf{a$_\mathbf{2}$} hBN nanoflake QEs (with g$^2(\tau = 0)$ values well below 0.5), and \textbf{a$_\mathbf{3}$} NV centers ensembles in diamond (with g$^2(\tau = 0)$ above 0.5). The dashed line presents the extracted lifetime of the emitter. \textbf{b$_\mathbf{i}$} The variation of the linear degree of polarization and \textbf{c$_\mathbf{i}$} axis of polarization measured with respect to time slices (time spent in the excited state) for different emitters. The dashed lines indicate the time-averaged visibility and polarization axis obtained by integrating over an extended time period (i.e., the results observed in Fig.\ \ref{fig-2}). For the irradiated hBN emitters and NV centres, measurements are performed with a pulsed excitation of 530 nm at 20 MHz and at 10 MHz repetition rates, respectively. Measurements for hBN nanoflake emitters are performed under 483 nm pulsed excitation at 10 MHz repetition rate.}
    \label{fig-3}
\end{figure*}

\indent We now turn our focus to the investigation of temporal polarization dynamics of the hBN (and in general of solid-state) quantum emitters. This study of the temporal emission polarization dynamics has been performed by recording the decay curve as a function of the rotation angle of the polarizer in the emission path. The algorithm to extract the relevant time-resolved polarization dynamics is described in Supplementary Section S6. As we are also interested in whether any observed effect is generic or only a sample-specific artefact, we also repeated this measurement for hBN nanoflakes and NV centers in nanodiamonds. The general optical characterization of these samples is shown in Supplementary Sections S7 and S8. Each decay measurement that corresponds to a different polarizer angle is then combined to obtain a polarization-resolved decay map. Fig.\ \ref{fig-3}a shows exemplary polarization-resolved decay maps from three different emitter types. Each map is then divided into time bins, and a generic cosine-squared function is fitted into each individual time bin to extract linear polarization visibility and the polarization axis (see Fig.\ \ref{fig-3}b and \ref{fig-3}c, respectively) as a function of the time that the charge carrier has spent in the excited state. In order to account for the instrument response function ($<$70 ps FWHM), the initial 120 ps of the data is omitted from the analysis. \\
\indent Interestingly, for all emitters investigated including the NV centers (we have used ensembles with 1-4 NV centers), we have observed a strong increase in the linear polarization visibility (see Fig.\ \ref{fig-3}b) during the first 1 to 3 nanoseconds. Accompanied by the visibility change, most of the emitters also show rotation in the polarization axis with respect to the decay time, while the polarization of others stays stable during this time scale. These effects could be caused by photo-induced modifications of the local charge environment around the emitter. In other words, the laser pulse excites other optically inactive nearby emitters or charge states which results in an induced electric field around the emitter \cite{Weston2018}. Such electric field fluctuations can temporarily shift the dipole axis of the emitters, resulting in a decrease in the observed polarization visibility. Alternatively, the used pulsed excitation laser in all our measurements carries very high peak laser power in every pulse. This could also induce local strain or structural fluctuation in the crystal for a short period of time, which then could also lead to temporal dynamics of the dipole polarization. Moreover, the pump laser illumination can also modify the electron occupation distribution around the defects, independent of local charges in the surrounding environment. If these were true mechanisms, then this should be laser-power dependent. Our power-resolved measurements (see Supplementary Section S9) have not indicated this; however, we are not necessarily ruling this out as the effect could be already saturated at all studied laser powers. In this case, when these undesired excitations or photo-induced modifications or strain relax on fast timescales, visibility and polarization axis reach their steady states. Importantly, the timescale of the relaxation is much longer than the laser pulse length ($<$70 ps FWHM) and varies from sample to sample. This could indicate different local charge distributions in the crystal environment or modifications in the crystal environment due to laser illumination.\\
\indent We also would like to remark on some of the discrepancies between polarization axis results on different samples. Observations show that both irradiated hBN emitters and NV centers (Fig.\ \ref{fig-3}c$_1$ and c$_3$, respectively) show a clear change in the polarization axis whereas emitters in hBN nanoflakes (Fig.\ \ref{fig-3}c$_2$) do not show such change. If these changes are indeed affected by the local environment of the emitter, we would expect sample preparation to play a crucial role in these different observations. The hBN nanoflakes are less clean in comparison to irradiated hBN quantum emitters and to NV centers in diamond in terms of local strains, charge states, or even nearby emitters. If the orientation of the dipole can be affected by its local environment, in a given time, we would expect such interactions to statistically average out each other in terms of dipole orientation. This would explain both the visibility change and orientation stability of the hBN nanoflakes. In the case of irradiated hBN samples, we would expect the local environment to be much cleaner and more ordered in comparison to nanoflakes. In the absence of such heavily random interactions, one might expect crystal axes to play a major role as a static force on the dipole orientation. Again, in a given time, the emitter might fluctuate its orientation in a preferred direction, but such fluctuations in the orientation would result in a decrease in the observation of visibility and orientation. The same discussion can also explain rotational changes in the orientation (Fig.\ \ref{fig-3}c$_1$, green line) since the rotational difference between an emitter and the (closest) crystal axis can be either $+$X or $-$X degrees.\\
\indent It is worth to mention that our experiments record the emission dipole orientation with respect to a fixed excitation dipole. The effect of excitation laser polarization dependence on the temporal dynamics is still unknown. A similar mechanism of temporal dynamics of polarization could exist for the excitation dipole, but it is challenging to distinguish from emission temporal polarization effect. Finally, we would like to draw attention to the results on the NV centers, which is a completely different type of quantum emitter, that surprisingly show the same behaviors as hBN emitters. The NV center is a very well-studied system and yet we are not aware of any such effect being reported. This raises the question of whether these observations are generic to the other emitter types in other materials, such as quantum dots \cite{PhysRevLett.116.020401}, quantum emitters in 2D transition metal dichalcogenides (TMDs) \cite{Tonndorf:15} and 3D crystals like diamond and silicon carbide.

\subsection*{Polarization dynamics with density functional theory}
\begin{figure*}[ht]
    \includegraphics[width = 1\textwidth]{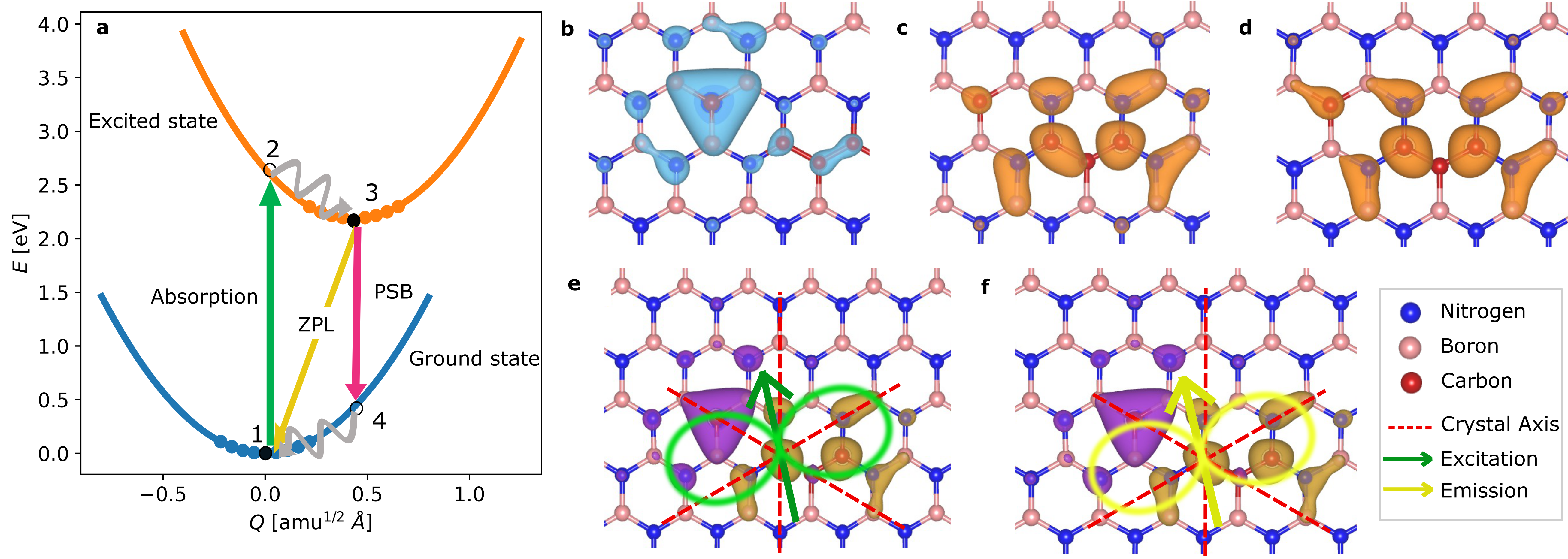}
    \caption{\textbf{a} Potential energy surface of a neutral-charged C$_{2}$C$_{2}$ defect without strain representing the complete excitation and emission process, consisting of the absorption (green line), the zero-phonon line (ZPL, yellow line, here 573 nm), and the phonon sideband (PSB, magenta line). \textbf{b}-\textbf{d} The probability density $|\psi|^2$ of electron occupations in the ground state at point 1 and excited states at points 2 and 3, respectively. \textbf{e} The charge difference between points 1 and 2 is shown by the isosurfaces, where the green arrow indicates the excitation dipole axis with light radiation in green shade. The excitation axis makes 11.1$^\circ$ relative to the crystal axis (red dashed line). \textbf{f} The charge difference between points 1 and 3 with the yellow arrow indicating the emission dipole axis with light radiation in yellow shade. The emission axis makes 12.1$^\circ$ relative to the crystal axis.}
    \label{fig:PES_DFT}
\end{figure*}
We now turn to theoretically modeling the observed effects with DFT. This section answers the questions: (i) Why are the excitation/emission dipoles misaligned? (ii) How do the dipoles specifically align with respect to the crystal axes? (iii) Can strain cause symmetry breaking? Lastly, (iv) can electric excess charges/defects cause temporal variations? We address these questions using spin-polarized density functional theory. Our DFT calculations use the HSE06 functional (see Methods), which provides reasonable accuracy for calculating the electronic band structures of hBN quantum emitters as verified by experiment compared to functionals from the generalized gradient approximation \cite{Reimers2018}. We have studied the most likely candidates, e.g., intrinsic defects and complexes involving oxygen and carbon impurities with neutral or $\pm$1 charge states. Carbon complexes could form during the SEM irradiation process, consistent with previous DFT calculations yielding 2 eV quantum emitters in hBN \cite{Cholsuk2022,Auburger2021,Mendelson2021}.\\
\indent The electronic transition of a defect, in theory, can be described by the Huang-Rhys model as shown in Fig.\ \ref{fig:PES_DFT}a, where the ground and excited states, depicted by the blue and orange curves, respectively, are responsible for the transition. Each state consists of vibrational modes, illustrated by the dots on the respective curve. In principle, a transition between any pair of ground and excited states is possible, resulting in the absorption and emission spectrum. Both consist of the zero-phonon line (ZPL) and a phonon sideband (PSB). We note that the transition dipole moment for the absorption can depend on the specific phonon mode (i.e., it is affected by the excitation laser wavelength relative to the ZPL) \cite{Jungwirth2017}. The excitation is the transition from points 1 to 2 in Fig.\ \ref{fig:PES_DFT}a. The system will relax to point 3 on ultrafast timescales (typically on the order of a few ps) through phonon scattering. From point 3 the emission can take place either directly to the ground state (point 1) or via another phonon mode (point 4). We have calculated both transition dipole moments for the ZPL (3$-$1 transition) and a (randomly chosen) phonon mode in the PSB (3$-$4 transition) and found only negligible difference in the relative angles (below 1$^\circ$, see Supplementary Section S10). In the experiment, one would see the average (i.e., the averaged dipole over ZPL and PSB which has a lower polarization visibility in case of a misalignment between both transition dipole moments). Hence, we restrict the following analysis to the ZPL transition (points 3 to 1). \\
\indent As these particular points originate from different electronic states and ionic relaxation configurations, their wave functions can be distinct. This is also consistent with previous DFT calculations \cite{Davidsson2020}. Fig.\ \ref{fig:PES_DFT}b-d show the probability densities ($|\psi|^2$) of the electron occupation of a defect (C$_{\text{B}}$C$_{\text{N}}$C$_{\text{B}}$C$_{\text{N}}$) corresponding to points 1 to 3, respectively. The differences between points 1 and 2, as well as 3 and 1, are shown in Fig.\ \ref{fig:PES_DFT}e and f, respectively. As the transition dipole moment is proportional to $\bra{\psi_{f}}\textbf{p}\ket{\psi_{i}}$ (see Eq.\ \ref{eq:dipole} in the Methods), where $i$ and $f$ denote the initial and final states of the transition, distinct wave functions can lead to different dipole moments and therefore also to an angle between the excitation and emission polarization. Hence, this answers our question (i). We are nevertheless not ruling out that even in our case the misaligned dipoles can be caused by additional intermediate states as has been reported before \cite{Jungwirth2017}.\\
\indent To identify the type of defect matching the experimental observations, we calculated properties of 126 native, carbon-, and oxygen-based defects with different charge-states (see Supplementary Section S10 and the attached data-set) and applied the following criteria to select most promising defect candidates: ZPL range, orientation of the absorption polarization axis relative to the crystal axis (in the range of 3.9$^\circ$ to 11.6$^\circ$), and the linear in-plane polarization visibility. We restrict these criteria to the absorption dipole, as our data is consistent with an excitation dipole that has a finite angle relative to the crystal axis. The interpretation of the emission data will be treated later in this work. The choices of oxygen and carbon for impurities are based on the fact that these have been suspected to be responsible for the 2 eV emission \cite{Li2022, Sajid2020-2}, as well as their natural chemical stability. Among the 126 studied defects, only 22 satisfy the range of the excitation polarization; however, most of these can be additionally ruled out due to the polarization visibility. In particular, many charged defects exhibit strong out-of-plane contributions. We found that the dipoles of charged-state defects are impacted by free excessive positive/negative charges and out-of-plane structural deformation, leading them to align perpendicularly to the crystal plane. This finding also confirms that the dipoles depend on the charge distribution and structural deformation. The remaining candidates can further be narrowed down when the ZPL is taken into account (see Supplementary Section S10). This essentially eliminates all defects except for the C$_{\text{B}}$C$_{\text{N}}$C$_{\text{B}}$C$_{\text{N}}$ defect complex. In the following, we will abbreviate this complex with C$_2$C$_2$. Due to having four carbon defects involved, it can exist naturally in several configurations (see Supplementary Section S10.4), which we denote with a number, i.e., C$_{2}$C$_{2}$-n. The most likely configurations are the neutral-charged C$_{2}$C$_{2}$-3 (which is shown in Fig.\ \ref{fig:PES_DFT}) and the neutral-charged C$_{2}$C$_{2}$-5 with ZPLs at 573 nm and 562 nm, respectively. With our current theoretical and experimental uncertainties, we cannot distinguish these cases definitely, and in principle, it would be even possible to have a mixture of both cases present in our sample. The atomic structures of these configurations are shown in Supplementary Section S10. Their excitation (emission) polarization axis aligns 11.1$^\circ$ (12.1$^\circ$) and 12.3$^\circ$ (13.7$^\circ$) relative to the nearest crystal axis, as marked by the arrows in Fig.\ \ref{fig:PES_DFT}e and f. This small difference between excitation and emission polarization axis is not consistent with the experimentally observed difference, which could be due to multiple factors causing the distortion of charge distribution in the experiments such as in-homogeneous distribution of strain in the flake or localized charge in the lattice due to the irradiation process. It is important to note that these structures undergo essentially only in-plane deformations during relaxation, making them inherit pure in-plane dipoles. Of course, the polarization can in principle be out-of-plane; nonetheless, our calculation indicates the in-plane polarization. We, therefore, propose the C$_{2}$C$_{2}$ defect to be responsible for our emission even though it exhibits notable variations in dipole angles compared to experimental observations. However, when considering overall properties (also ZPL, polarization visibility, etc.), the C$_{2}$C$_{2}$ defect emerges as the most favorable emitter compared to all other studied defects (see Supplementary Section S10). This also provides the answer to the question (ii).\\
\indent What our model so far could not explain is the symmetry breaking, i.e., the center of the groups in the emission axis not being centered around the crystal axes and also not spaced by 60$^\circ$ apart as well as the large distribution of e.g.\ angle difference (excitation minus emission). We speculate that this could be caused by strain. One has to distinguish two cases here: global and local strains. We know the former is not significant, as otherwise the SHG pattern would be skewed. Nevertheless, there could be a significant amount of local strain around the irradiated spots. We can model whether this is possible qualitatively using DFT as well. We have (theoretically) applied bi-axial strain in the range of $\pm$1\% to the lattice and monitored the dipole orientations. For vacancy-based defects, the shifts can amount to more than 4$^\circ$, while for the C$_2$C$_2$-3 for example this remains below 0.5$^\circ$ in the investigated strain range (see Supplementary Section S10). Moreover, we also observe that the excitation and emission dipoles are affected differently by local strain, which could account in part for symmetry breaking, hence unequal shifts from the crystal axis. We can therefore answer the question (iii) only qualitatively in part and attribute either large local strain around the irradiated spots or other (so far unknown) local modifications in the crystal environment as the cause of the symmetry breaking. The quantitative description is beyond the scope of this work and will be carried out by us in a future study.\\
\indent Finally, this leaves the question of polarization (iv) to be answered. To model the temporal variations observed in Fig.\ \ref{fig-3}, we applied an external electric field (up to 0.7 V/\AA) to mimic the redistribution of the local charge environment around the defects. The results (see Supplementary Section S10) indicate that for all defects the in-plane dipoles turn out to have a high out-of-plane contribution with high intensity of an out-of-plane electric field. We only consider the limit of weak electric fields that only have a minor impact on the photophysical properties (i.e., adiabatic changes only and no jumps). In this limit, the reduction in visibility can be substantial ($>$\,20\%) together with a polarization axis of rotation $>$\,5$^\circ$. This confirms (again qualitatively) that the dipoles are sensitive to the local charge distribution, and this could explain why the polarization visibility increases when reaching the steady state as depicted in Fig.\ \ref{fig-3}c.

\section{Conclusion}
The present work demonstrates an in-depth study of the polarization dynamics of identical yellow single photon emitters in hBN fabricated using a standard electron beam microscope. Our findings indicate a correlation between the excitation and emission axes and the crystallographic hBN crystal axis. While the excitation polarization bunches around the crystal axes, we found that the emission polarization bunches in between the crystal axes. As the latter groups are not separated by 60$^\circ$, we suggest that local strain could cause the symmetry breaking. The correlation of crystal axes with dipole polarizations of quantum emitters can, in principle, be used to identify the emitter in question when compared with predictions calculated with DFT. The direct identification of hBN quantum emitters has been shown to be technically difficult in the past. While in this work we have a large variation in the observed polarization angles to undoubtedly identify the emitter, we can narrow down potential defect candidates. When this is done together with other photophysical properties such as emission spectrum, this can lead to a convincing case for the proposed atomic structure to be responsible for the 2 eV quantum emitter in hBN \cite{cholsuk2023-comprehensive}. This could also provide an approach to address the atomic structures of fluorescent defects in other materials systems, such as TMDs \cite{Tonndorf:15,nnano.2015.60,nnano.2015.75}, silicon carbide \cite{Lohrmann2015}, and silicon \cite{Redjem.2023}.\\
\indent We have also investigated the temporal dynamics of the polarization of single photons generated from defects in irradiated and nanoflake hBN as well as the negatively-charged nitrogen vacancy center in diamond. A higher degree of emission polarization is observed for the carriers that stay longer in the excited state. We speculate that this effect could be due to the local electric field induced by the excess charges that are excited with the laser pulse. This effect can also be explained in terms of photo-induced strain or modifications in the local charge distribution under pulsed excitation. To complement our experimental observations, we provide DFT calculations. Nevertheless, the spin Hamiltonian simulation is further required for studying the polarization dynamics. We believe that the observed temporal change of polarization in various solid-state quantum emitter systems is critical to reaching the ideal performance of these emitters for several applications, such as to generate the Fourier-transform limited photons \cite{Komza.Sipahigil.2022, Fournier.Delteil.2022} or to achieve lower quantum bit error rate in quantum key distribution systems \cite{Caglar, Gao.Heindel.2023}. It might even be an important step toward achieving indistinguishable single photons from a room-temperature solid-state quantum light source when coupled with resonant structures \cite{10.1021/acsphotonics.9b00314}.

\section*{Methods}
\subsection*{Emitter fabrication} 
A multilayer hBN flake was exfoliated from a bulk crystal (HQ Graphene) using the scotch tape method onto a visco-elastic polymer sheet (Polydimethylsiloxane) purchased from Gel-Pak (WF-40-X4). The exfoliated flakes were examined under a bright field optical microscope to identify a suitable thin flake based on the optical contrast. Afterward, the flake was transferred onto a grid-patterned Si/SiO$_2$ substrate with a 298 nm thermal oxide layer. This grid was fabricated using electron beam lithography and a metal lift-off process, which allowed us to easily navigate on the substrate.\\
\indent The nanoflake emitters were obtained in solution with a concentration of 5.5 mg/l from Graphene Supermarket. The number of atomic layers per flake varies between one and five with a typical flake diameter ranging from 50 to 200 nm. Approximately 10 $\mu$l of the solution was drop-cast onto a Si/SiO$_2$ substrate with a 300 nm oxide layer and dried under ambient conditions. No further post-processing such as high-temperature annealing was carried out.\\
\indent The nanodiamonds were prepared by drop-casting commercially available nanodiamonds solution (Adamas Nanotechnologies, 40 nm Carboxylated Red FND 1-4 NV per particle) on standard glass substrates and stored at ambient conditions overnight to dry.

\subsection*{Emitter irradiation}
The emitter array was produced using a scanning electron microscope (Helios NanoLab G3). The electron beam is accelerated at 3 kV with an electron current of 25 pA. These settings were used for beam alignment, imaging, and the actual irradiation. The imaging of the flake was carried out with an electron fluence of $1.4\times 10^{13}$ cm$^{-2}$. To fabricate the emitters, a high electron flux was pointed for a dwell time of 10 s (fluence $7.7\times 10^{17}$ cm$^{-2}$) onto pre-defined spots on a suitable flake. 

\subsection*{Optical characterization}
The optical investigation of the hBN emitter array and the NV samples was carried out using a commercial fluorescence lifetime imaging microscope (PicoQuant MicroTime 200) with a 530 nm pulsed laser at 20 MHz repetition rate and a pulse length below 80 ps (FWHM). For the NV samples, we have reduced the repetition rate to 5 MHz (to account for their higher excited state lifetime). Unless stated otherwise, the excitation power for all measurements was around 50 $\mu W$ (peak power $\geq$ 10 mW). For the PL mapping using a scanning stage, a dwell time of 5 ms per pixel is used. The laser is circularly polarized with a quarter wave-plate and then linearly using a nanoparticle film polarizer on a motorized mount. The PL signal is collected using a 100$\times$ dry immersion objective with a high numerical aperture (NA) of 0.9 and a working distance of 0.3 mm. In the detection path, we have inserted a long-pass filter to suppress the excitation laser and another motorized nanoparticle film polarizer. The photons are detected by two single photon avalanche diodes from Micro Photon Devices or a high-resolution spectrometer. The assembly of the SPADs in both arms of a 50:50 beam splitter enables us to measure the second-order correlation function. The data analysis of the correlation function as well as the lifetime measurements is performed with the built-in software (that also takes the instrument response function into account by convoluting the initial fit function with the measured instrument response function of $<$70 ps FWHM and then using the resulting function to fit the data). The spectral data was obtained with an acquisition time of 1\, min per emitter.\\
\indent The optical properties of hBN nanoflake emitters were studied using a custom-built confocal microscope setup. The setup comprises various pulsed lasers (Advanced Laser Diode Systems, Pilas) with wavelengths of 405, 483, and 637 nm and pulse lengths below 50 ps, an objective with an NA of 0.75 and 50$\times$ magnification, a spectrometer with a resolution of 0.03 nm (Andor, Shamrock 750) together with a CCD camera (Andor, Newton), and four SPADs (ID Quantique, 2$\times$ ID120, and 2$\times$ ID100) located at the detection ports. The excitation power used in all measurements was around 100 $\mu W$, unless stated otherwise, which is below the saturation power of the emitters. A combination of long-pass and notch filters is used at the detection port to filter out the excitation laser, while various bandpass filters are employed to selectively filter out the PL emission from different emitters. Time-correlated single photon counting was performed using a time-tagger module (Roithner LaserTechnik TTM8000) with a resolution of 41 ps to record the event times. To prevent intensity variations dependent on acquisition, a fixed acquisition time of 1 minute is used for each polarizer angle in the temporal polarization measurements.

During all temporal polarization measurements, signals are either filtered by spectral filters or a spectrometer to exclude the excitation laser that is reflected from the sample surface. Additionally, we also monitor the anti-bunching measurements for the emitters. 

\subsection*{SHG Characterization}
A pulsed Ti:Sapphire laser (Coherent Verdi \& Chameleon) was used as the pump source for the SHG measurements. The laser wavelength was set to 800 nm and has a pulse duration of 200 fs (estimated at the sample position) at a repetition rate of 76 MHz. The power is controlled by a half-wave plate (HWP) and a polarizing beam splitter (PBS). The pump laser polarization was controlled by a motorized HWP and is coupled to the sample with a beam splitter (BS) and a 50$\times$, 0.55 NA objective (Zeiss LD EC Epiplan-Neofluar). The reflection at 800 nm (pump) and transmission at 400 nm (second-harmonic) of this BS is similar for both polarization components. The sample was mounted on an XY-motorized stage for position control and the objective was on a Z-axis motorized stage to control the focus. The generated second-harmonic light is coupled from the sample in reflection using the same objective and separated from the pump by the BS. A motor-controlled polarizer enables polarization analysis of the SHG process, and a spectral filter (BG39) is used to remove excess pump light before detection with a thermoelectrically cooled CCD (Andor Zyla 4.2P). The average power of the Ti:Sapphire laser was set to 20 mW before the objective, which is sufficient to observe the SHG signal and low enough to not damage the hBN flake. The polarization scans involve rotating the HWP and Polarizer in either a parallel configuration $(\theta_{\mathrm{HWP}}=\frac{\theta_{\mathrm{pol}}}{2})$ or in the perpendicular configuration $(\theta_{\mathrm{HWP}}=\frac{\theta_{\mathrm{pol}} + \pi}{2})$. At each polarization setting, the CCD intensity was integrated for 10 seconds, and finally, a background subtraction with the pump laser turned off was used to improve the contrast of the data.

\subsection*{DFT calculations}
All spin-polarized DFT calculations were performed using the Vienna Ab initio Simulation Package (VASP) with a plane wave basis set \cite{vasp1,vasp2} and the projector augmented wave (PAW) as the pseudopotentials \cite{paw,paw2}. The sizes of the vacuum layer and supercell were optimized until the hBN band structure remains unchanged, which yields a 15 \AA $ $ vacuum layer and a $7\times 7\times 1$ supercell size containing 98 atoms. The HSE06 functional was employed for all calculations as it was known to yield more reliable results with the experiment than the generalized gradient approximation \cite{Reimers2018,Sajid2020}. The single $\Gamma$-point calculation was implemented to relax the structures with only internal coordinates allowed until the force is lower than 0.01 eV/\AA. All geometry relaxations were performed with an energy cutoff at 500 eV and the total energy convergence with the accuracy of 10$^{-4}$ eV. For the excited-state calculations, we used the $\Delta$SCF method to constrain the electron occupation in the excited-state configuration. The transition dipole moment (TDM) \cite{Davidsson2020} is expressed by
\begin{equation}
    \textbf{$\mu$} = \frac{i\hbar}{(E_{f} - E_{i})m}\bra{\psi_{f}}\textbf{p}\ket{\psi_{i}},
    \label{eq:dipole}
\end{equation}
where $E_{i}$ and $E_{f}$ are the eigenvalues of the initial and final orbitals, accordingly, $m$ is the mass of an electron, and \textbf{p} is the momentum operator. All other computational details can be found in Supplementary Section S10. Note that as the excitation/emission polarization axes are perpendicular to the dipole axes, hence, we projected and rotated the calculated dipole axes to be consistent with the experiments. To extract the wave function, the PyVaspwfc Python code \cite{PyVaspwfc} and the modified version \cite{Davidsson2020} were implemented.  Finally, we applied an out-of-plane external electric field along with the dipole correction to prevent the error from the periodic condition for an electric field simulation. To investigate whether the strain changes the dipole orientation, the bi-axial strain was applied. Note that for both electric field and strain calculations, we set the force to be 0.02 eV/\AA$ $ to reduce computational time.

\section*{Data availability}\label{dataset}
All data from this work is available from the authors upon reasonable request. 

\section*{Supporting Information Available}
\begin{itemize}
    \item (1)  General photon-physical properties of the yellow emitters, (2) Polarization dynamics data acquisition (3) Polarization dynamics data analysis (4) Second-harmonic generation measurement (5) Misalignment between excitation, emission axis and crystal axis (6) Temporal polarization dynamics (7) hBN nanoflake quantum emitters (8) NV centers in diamond (9) Power-dependent temporal dynamics of polarization (10) DFT calculations. 
    \item The complete dataset for defect candidates, as determined through DFT, can be accessed through the provided link: \url{https://doi.org/10.5281/zenodo.10288562}
\end{itemize}

\section*{Notes}
The authors declare no competing financial interest.

\section*{Author contributions}
A.K., C.S., and A.Z. prepared the samples. A.K., C.S., T.M., S.P., and Y.O. performed the optical characterization. A.K. performed the irradiation on the hBN flake. R.J.C., G.S., and R.G. carried out the SHG measurements. C.C. and S.S. performed the DFT calculation. A.K., C.S., and C.C. analyzed the data. T.V. and S.A. conceived and supervised the work. 

\begin{acknowledgments}
This work was funded by the Federal Ministry of Education and Research (BMBF) under grant number 13N16292 and the Deutsche Forschungsgemeinschaft (DFG, German Research Foundation) - Projektnummer 445275953. The authors acknowledge support by the German Space Agency DLR with funds provided by the Federal Ministry for Economic Affairs and Climate Action BMWK under grant numbers 50WM2165 (QUICK3) and 50RP2200 (QuVeKS). The major instrumentation used in this work was funded by the Free State of Thuringia via the projects 2015 FOR 0005 (ACP-FIB) and 2017 IZN 0012 (InQuoSens). C.C. acknowledges a Development and Promotion of Science and Technology Talents Project (DPST) scholarship by the Royal Thai Government. The computational experiments were supported by resources of the Friedrich Schiller University Jena supported in part by DFG grants INST 275/334-1 FUGG and INST 275/363-1 FUGG. This work was supported by the Scientific and Technological Research Council of Turkey (TUBITAK) under project numbers 117F495 and 118E994. S.A. acknowledges the support by the Turkish Academy of Sciences (T\"UBA-GEBIP; The Young Scientist Award Program) and the Science Academy of Turkey (BAGEP; The Young Scientist Award Program). S.S. acknowledges funding support by Mahidol University (Fundamental Fund: fiscal year 2023 by National Science Research and Innovation Fund (NSRF)) and from the NSRF via the Program Management Unit for Human Resources \& Institutional Development, Research and Innovation (grant number B05F650024). R.J.C. acknowledges funding from the Swiss National Science Foundation under the Ambizione Fellowship Program (Project Number 208707). R.G. acknowledges funding from the European Space Agency (OSIP HEIDI Project Number 4000137426). We are grateful to Joel Davidsson for the source code of transition dipole moments for two wave functions.
\end{acknowledgments}


\bibliography{main}

\providecommand{\noopsort}[1]{}\providecommand{\singleletter}[1]{#1}%
\begin{thebibliography}{60}%
\makeatletter
\providecommand \@ifxundefined [1]{%
 \@ifx{#1\undefined}
}%
\providecommand \@ifnum [1]{%
 \ifnum #1\expandafter \@firstoftwo
 \else \expandafter \@secondoftwo
 \fi
}%
\providecommand \@ifx [1]{%
 \ifx #1\expandafter \@firstoftwo
 \else \expandafter \@secondoftwo
 \fi
}%
\providecommand \natexlab [1]{#1}%
\providecommand \enquote  [1]{``#1''}%
\providecommand \bibnamefont  [1]{#1}%
\providecommand \bibfnamefont [1]{#1}%
\providecommand \citenamefont [1]{#1}%
\providecommand \href@noop [0]{\@secondoftwo}%
\providecommand \href [0]{\begingroup \@sanitize@url \@href}%
\providecommand \@href[1]{\@@startlink{#1}\@@href}%
\providecommand \@@href[1]{\endgroup#1\@@endlink}%
\providecommand \@sanitize@url [0]{\catcode `\\12\catcode `\$12\catcode
  `\&12\catcode `\#12\catcode `\^12\catcode `\_12\catcode `\%12\relax}%
\providecommand \@@startlink[1]{}%
\providecommand \@@endlink[0]{}%
\providecommand \url  [0]{\begingroup\@sanitize@url \@url }%
\providecommand \@url [1]{\endgroup\@href {#1}{\urlprefix }}%
\providecommand \urlprefix  [0]{URL }%
\providecommand \Eprint [0]{\href }%
\providecommand \doibase [0]{https://doi.org/}%
\providecommand \selectlanguage [0]{\@gobble}%
\providecommand \bibinfo  [0]{\@secondoftwo}%
\providecommand \bibfield  [0]{\@secondoftwo}%
\providecommand \translation [1]{[#1]}%
\providecommand \BibitemOpen [0]{}%
\providecommand \bibitemStop [0]{}%
\providecommand \bibitemNoStop [0]{.\EOS\space}%
\providecommand \EOS [0]{\spacefactor3000\relax}%
\providecommand \BibitemShut  [1]{\csname bibitem#1\endcsname}%
\let\auto@bib@innerbib\@empty
\bibitem [{\citenamefont {Aharonovich}\ \emph {et~al.}(2016)\citenamefont
  {Aharonovich}, \citenamefont {Englund},\ and\ \citenamefont
  {Toth}}]{10.1038/nphoton.2016.186}%
  \BibitemOpen
  \bibfield  {author} {\bibinfo {author} {\bibfnamefont {I.}~\bibnamefont
  {Aharonovich}}, \bibinfo {author} {\bibfnamefont {D.}~\bibnamefont
  {Englund}},\ and\ \bibinfo {author} {\bibfnamefont {M.}~\bibnamefont
  {Toth}},\ }\bibfield  {title} {\bibinfo {title} {Solid-state single-photon
  emitters},\ }\href {https://doi.org/10.1038/nphoton.2016.186} {\bibfield
  {journal} {\bibinfo  {journal} {Nat. Photon.}\ }\textbf {\bibinfo {volume}
  {10}},\ \bibinfo {pages} {631} (\bibinfo {year} {2016})}\BibitemShut
  {NoStop}%
\bibitem [{\citenamefont {Atat{\"u}re}\ \emph {et~al.}(2018)\citenamefont
  {Atat{\"u}re}, \citenamefont {Englund}, \citenamefont {Vamivakas},
  \citenamefont {Lee},\ and\ \citenamefont {Wrachtrup}}]{Atature2018}%
  \BibitemOpen
  \bibfield  {author} {\bibinfo {author} {\bibfnamefont {M.}~\bibnamefont
  {Atat{\"u}re}}, \bibinfo {author} {\bibfnamefont {D.}~\bibnamefont
  {Englund}}, \bibinfo {author} {\bibfnamefont {N.}~\bibnamefont {Vamivakas}},
  \bibinfo {author} {\bibfnamefont {S.-Y.}\ \bibnamefont {Lee}},\ and\ \bibinfo
  {author} {\bibfnamefont {J.}~\bibnamefont {Wrachtrup}},\ }\bibfield  {title}
  {\bibinfo {title} {Material platforms for spin-based photonic quantum
  technologies},\ }\href {https://doi.org/10.1038/s41578-018-0008-9} {\bibfield
   {journal} {\bibinfo  {journal} {Nat. Rev. Mater.}\ }\textbf {\bibinfo
  {volume} {3}},\ \bibinfo {pages} {38} (\bibinfo {year} {2018})}\BibitemShut
  {NoStop}%
\bibitem [{\citenamefont {O'Brien}(2007)}]{Quantumcomp}%
  \BibitemOpen
  \bibfield  {author} {\bibinfo {author} {\bibfnamefont {J.~L.}\ \bibnamefont
  {O'Brien}},\ }\bibfield  {title} {\bibinfo {title} {Optical quantum
  computing},\ }\href {https://doi.org/10.1126/science.1142892} {\bibfield
  {journal} {\bibinfo  {journal} {Science}\ }\textbf {\bibinfo {volume}
  {318}},\ \bibinfo {pages} {1567} (\bibinfo {year} {2007})}\BibitemShut
  {NoStop}%
\bibitem [{\citenamefont {Lo}\ \emph {et~al.}(2014)\citenamefont {Lo},
  \citenamefont {Curty},\ and\ \citenamefont {Tamaki}}]{Lo2014}%
  \BibitemOpen
  \bibfield  {author} {\bibinfo {author} {\bibfnamefont {H.-K.}\ \bibnamefont
  {Lo}}, \bibinfo {author} {\bibfnamefont {M.}~\bibnamefont {Curty}},\ and\
  \bibinfo {author} {\bibfnamefont {K.}~\bibnamefont {Tamaki}},\ }\bibfield
  {title} {\bibinfo {title} {Secure quantum key distribution},\ }\href
  {https://doi.org/10.1038/nphoton.2014.149} {\bibfield  {journal} {\bibinfo
  {journal} {Nat. Photon.}\ }\textbf {\bibinfo {volume} {8}},\ \bibinfo {pages}
  {595} (\bibinfo {year} {2014})}\BibitemShut {NoStop}%
\bibitem [{\citenamefont {Degen}\ \emph {et~al.}(2017)\citenamefont {Degen},
  \citenamefont {Reinhard},\ and\ \citenamefont {Cappellaro}}]{Degen2017}%
  \BibitemOpen
  \bibfield  {author} {\bibinfo {author} {\bibfnamefont {C.}~\bibnamefont
  {Degen}}, \bibinfo {author} {\bibfnamefont {F.}~\bibnamefont {Reinhard}},\
  and\ \bibinfo {author} {\bibfnamefont {P.}~\bibnamefont {Cappellaro}},\
  }\bibfield  {title} {\bibinfo {title} {Quantum sensing},\ }\href
  {https://doi.org/10.1103/RevModPhys.89.035002} {\bibfield  {journal}
  {\bibinfo  {journal} {Rev. Mod. Phys.}\ }\textbf {\bibinfo {volume} {89}},\
  \bibinfo {pages} {035002} (\bibinfo {year} {2017})}\BibitemShut {NoStop}%
\bibitem [{\citenamefont {Vogl}\ \emph {et~al.}(2021)\citenamefont {Vogl},
  \citenamefont {Knopf}, \citenamefont {Weissflog}, \citenamefont {Lam},\ and\
  \citenamefont {Eilenberger}}]{PhysRevResearch.3.013296}%
  \BibitemOpen
  \bibfield  {author} {\bibinfo {author} {\bibfnamefont {T.}~\bibnamefont
  {Vogl}}, \bibinfo {author} {\bibfnamefont {H.}~\bibnamefont {Knopf}},
  \bibinfo {author} {\bibfnamefont {M.}~\bibnamefont {Weissflog}}, \bibinfo
  {author} {\bibfnamefont {P.~K.}\ \bibnamefont {Lam}},\ and\ \bibinfo {author}
  {\bibfnamefont {F.}~\bibnamefont {Eilenberger}},\ }\bibfield  {title}
  {\bibinfo {title} {Sensitive single-photon test of extended quantum theory
  with two-dimensional hexagonal boron nitride},\ }\href
  {https://link.aps.org/doi/10.1103/PhysRevResearch.3.013296} {\bibfield
  {journal} {\bibinfo  {journal} {Phys. Rev. Res.}\ }\textbf {\bibinfo {volume}
  {3}},\ \bibinfo {pages} {013296} (\bibinfo {year} {2021})}\BibitemShut
  {NoStop}%
\bibitem [{\citenamefont {Tran}\ \emph {et~al.}(2016)\citenamefont {Tran},
  \citenamefont {Bray}, \citenamefont {Ford}, \citenamefont {Toth},\ and\
  \citenamefont {Aharonovich}}]{nnano.2015.242}%
  \BibitemOpen
  \bibfield  {author} {\bibinfo {author} {\bibfnamefont {T.~T.}\ \bibnamefont
  {Tran}}, \bibinfo {author} {\bibfnamefont {K.}~\bibnamefont {Bray}}, \bibinfo
  {author} {\bibfnamefont {M.~J.}\ \bibnamefont {Ford}}, \bibinfo {author}
  {\bibfnamefont {M.}~\bibnamefont {Toth}},\ and\ \bibinfo {author}
  {\bibfnamefont {I.}~\bibnamefont {Aharonovich}},\ }\bibfield  {title}
  {\bibinfo {title} {Quantum emission from hexagonal boron nitride
  monolayers},\ }\href {http://dx.doi.org/nnano.2015.242} {\bibfield  {journal}
  {\bibinfo  {journal} {Nat. Nanotechnol.}\ }\textbf {\bibinfo {volume} {11}},\
  \bibinfo {pages} {37} (\bibinfo {year} {2016})}\BibitemShut {NoStop}%
\bibitem [{\citenamefont {Bradac}\ \emph {et~al.}(2019)\citenamefont {Bradac},
  \citenamefont {Gao}, \citenamefont {Forneris}, \citenamefont {Trusheim},\
  and\ \citenamefont {Aharonovich}}]{Bradac2019}%
  \BibitemOpen
  \bibfield  {author} {\bibinfo {author} {\bibfnamefont {C.}~\bibnamefont
  {Bradac}}, \bibinfo {author} {\bibfnamefont {W.}~\bibnamefont {Gao}},
  \bibinfo {author} {\bibfnamefont {J.}~\bibnamefont {Forneris}}, \bibinfo
  {author} {\bibfnamefont {M.~E.}\ \bibnamefont {Trusheim}},\ and\ \bibinfo
  {author} {\bibfnamefont {I.}~\bibnamefont {Aharonovich}},\ }\bibfield
  {title} {\bibinfo {title} {Quantum nanophotonics with group iv defects in
  diamond},\ }\href {https://doi.org/10.1038/s41467-019-13332-w} {\bibfield
  {journal} {\bibinfo  {journal} {Nat. Commun.}\ }\textbf {\bibinfo {volume}
  {10}},\ \bibinfo {pages} {5625} (\bibinfo {year} {2019})}\BibitemShut
  {NoStop}%
\bibitem [{\citenamefont {Senichev}\ \emph {et~al.}(2021)\citenamefont
  {Senichev}, \citenamefont {Martin}, \citenamefont {Peana}, \citenamefont
  {Sychev}, \citenamefont {Xu}, \citenamefont {Lagutchev}, \citenamefont
  {Boltasseva},\ and\ \citenamefont {Shalaev}}]{doi:10.1126/sciadv.abj0627}%
  \BibitemOpen
  \bibfield  {author} {\bibinfo {author} {\bibfnamefont {A.}~\bibnamefont
  {Senichev}}, \bibinfo {author} {\bibfnamefont {Z.~O.}\ \bibnamefont
  {Martin}}, \bibinfo {author} {\bibfnamefont {S.}~\bibnamefont {Peana}},
  \bibinfo {author} {\bibfnamefont {D.}~\bibnamefont {Sychev}}, \bibinfo
  {author} {\bibfnamefont {X.}~\bibnamefont {Xu}}, \bibinfo {author}
  {\bibfnamefont {A.~S.}\ \bibnamefont {Lagutchev}}, \bibinfo {author}
  {\bibfnamefont {A.}~\bibnamefont {Boltasseva}},\ and\ \bibinfo {author}
  {\bibfnamefont {V.~M.}\ \bibnamefont {Shalaev}},\ }\bibfield  {title}
  {\bibinfo {title} {Room-temperature single-photon emitters in silicon
  nitride},\ }\href {https://doi.org/10.1126/sciadv.abj0627} {\bibfield
  {journal} {\bibinfo  {journal} {Sci. Adv.}\ }\textbf {\bibinfo {volume}
  {7}},\ \bibinfo {pages} {eabj0627} (\bibinfo {year} {2021})}\BibitemShut
  {NoStop}%
\bibitem [{\citenamefont {Morfa}\ \emph {et~al.}(2012)\citenamefont {Morfa},
  \citenamefont {Gibson}, \citenamefont {Karg}, \citenamefont {Karle},
  \citenamefont {Greentree}, \citenamefont {Mulvaney},\ and\ \citenamefont
  {Tomljenovic-Hanic}}]{doi:10.1021/nl204010e}%
  \BibitemOpen
  \bibfield  {author} {\bibinfo {author} {\bibfnamefont {A.~J.}\ \bibnamefont
  {Morfa}}, \bibinfo {author} {\bibfnamefont {B.~C.}\ \bibnamefont {Gibson}},
  \bibinfo {author} {\bibfnamefont {M.}~\bibnamefont {Karg}}, \bibinfo {author}
  {\bibfnamefont {T.~J.}\ \bibnamefont {Karle}}, \bibinfo {author}
  {\bibfnamefont {A.~D.}\ \bibnamefont {Greentree}}, \bibinfo {author}
  {\bibfnamefont {P.}~\bibnamefont {Mulvaney}},\ and\ \bibinfo {author}
  {\bibfnamefont {S.}~\bibnamefont {Tomljenovic-Hanic}},\ }\bibfield  {title}
  {\bibinfo {title} {Single-photon emission and quantum characterization of
  zinc oxide defects},\ }\href {https://doi.org/10.1021/nl204010e} {\bibfield
  {journal} {\bibinfo  {journal} {Nano Lett.}\ }\textbf {\bibinfo {volume}
  {12}},\ \bibinfo {pages} {949} (\bibinfo {year} {2012})}\BibitemShut
  {NoStop}%
\bibitem [{\citenamefont {Vogl}\ \emph
  {et~al.}(2019{\natexlab{a}})\citenamefont {Vogl}, \citenamefont {Doherty},
  \citenamefont {Buchler}, \citenamefont {Lu},\ and\ \citenamefont
  {Lam}}]{C9NR04269E}%
  \BibitemOpen
  \bibfield  {author} {\bibinfo {author} {\bibfnamefont {T.}~\bibnamefont
  {Vogl}}, \bibinfo {author} {\bibfnamefont {M.~W.}\ \bibnamefont {Doherty}},
  \bibinfo {author} {\bibfnamefont {B.~C.}\ \bibnamefont {Buchler}}, \bibinfo
  {author} {\bibfnamefont {Y.}~\bibnamefont {Lu}},\ and\ \bibinfo {author}
  {\bibfnamefont {P.~K.}\ \bibnamefont {Lam}},\ }\bibfield  {title} {\bibinfo
  {title} {Atomic localization of quantum emitters in multilayer hexagonal
  boron nitride},\ }\href {https://doi.org/10.1039/C9NR04269E} {\bibfield
  {journal} {\bibinfo  {journal} {Nanoscale}\ }\textbf {\bibinfo {volume}
  {11}},\ \bibinfo {pages} {14362} (\bibinfo {year}
  {2019}{\natexlab{a}})}\BibitemShut {NoStop}%
\bibitem [{\citenamefont {Doherty}\ \emph {et~al.}(2013)\citenamefont
  {Doherty}, \citenamefont {Manson}, \citenamefont {Delaney}, \citenamefont
  {Jelezko}, \citenamefont {Wrachtrup},\ and\ \citenamefont
  {Hollenberg}}]{DOHERTY20131}%
  \BibitemOpen
  \bibfield  {author} {\bibinfo {author} {\bibfnamefont {M.~W.}\ \bibnamefont
  {Doherty}}, \bibinfo {author} {\bibfnamefont {N.~B.}\ \bibnamefont {Manson}},
  \bibinfo {author} {\bibfnamefont {P.}~\bibnamefont {Delaney}}, \bibinfo
  {author} {\bibfnamefont {F.}~\bibnamefont {Jelezko}}, \bibinfo {author}
  {\bibfnamefont {J.}~\bibnamefont {Wrachtrup}},\ and\ \bibinfo {author}
  {\bibfnamefont {L.~C.}\ \bibnamefont {Hollenberg}},\ }\bibfield  {title}
  {\bibinfo {title} {The nitrogen-vacancy colour centre in diamond},\ }\href
  {https://doi.org/https://doi.org/10.1016/j.physrep.2013.02.001} {\bibfield
  {journal} {\bibinfo  {journal} {Phys. Rep.}\ }\textbf {\bibinfo {volume}
  {528}},\ \bibinfo {pages} {1} (\bibinfo {year} {2013})}\BibitemShut {NoStop}%
\bibitem [{\citenamefont {Choi}\ \emph {et~al.}(2016)\citenamefont {Choi},
  \citenamefont {Tran}, \citenamefont {Elbadawi}, \citenamefont {Lobo},
  \citenamefont {Wang}, \citenamefont {Juodkazis}, \citenamefont {Seniutinas},
  \citenamefont {Toth},\ and\ \citenamefont
  {Aharonovich}}]{10.1021/acsami.6b09875}%
  \BibitemOpen
  \bibfield  {author} {\bibinfo {author} {\bibfnamefont {S.}~\bibnamefont
  {Choi}}, \bibinfo {author} {\bibfnamefont {T.~T.}\ \bibnamefont {Tran}},
  \bibinfo {author} {\bibfnamefont {C.}~\bibnamefont {Elbadawi}}, \bibinfo
  {author} {\bibfnamefont {C.}~\bibnamefont {Lobo}}, \bibinfo {author}
  {\bibfnamefont {X.}~\bibnamefont {Wang}}, \bibinfo {author} {\bibfnamefont
  {S.}~\bibnamefont {Juodkazis}}, \bibinfo {author} {\bibfnamefont
  {G.}~\bibnamefont {Seniutinas}}, \bibinfo {author} {\bibfnamefont
  {M.}~\bibnamefont {Toth}},\ and\ \bibinfo {author} {\bibfnamefont
  {I.}~\bibnamefont {Aharonovich}},\ }\bibfield  {title} {\bibinfo {title}
  {Engineering and localization of quantum emitters in large hexagonal boron
  nitride layers},\ }\href {https://doi.org/10.1021/acsami.6b09875} {\bibfield
  {journal} {\bibinfo  {journal} {ACS Appl. Mater. Interfaces}\ }\textbf
  {\bibinfo {volume} {8}},\ \bibinfo {pages} {29642} (\bibinfo {year}
  {2016})}\BibitemShut {NoStop}%
\bibitem [{\citenamefont {Vogl}\ \emph {et~al.}(2018)\citenamefont {Vogl},
  \citenamefont {Campbell}, \citenamefont {Buchler}, \citenamefont {Lu},\ and\
  \citenamefont {Lam}}]{10.1021/acsphotonics.8b00127}%
  \BibitemOpen
  \bibfield  {author} {\bibinfo {author} {\bibfnamefont {T.}~\bibnamefont
  {Vogl}}, \bibinfo {author} {\bibfnamefont {G.}~\bibnamefont {Campbell}},
  \bibinfo {author} {\bibfnamefont {B.~C.}\ \bibnamefont {Buchler}}, \bibinfo
  {author} {\bibfnamefont {Y.}~\bibnamefont {Lu}},\ and\ \bibinfo {author}
  {\bibfnamefont {P.~K.}\ \bibnamefont {Lam}},\ }\bibfield  {title} {\bibinfo
  {title} {Fabrication and deterministic transfer of high-quality quantum
  emitters in hexagonal boron nitride},\ }\href
  {https://doi.org/10.1021/acsphotonics.8b00127} {\bibfield  {journal}
  {\bibinfo  {journal} {ACS Photonics}\ }\textbf {\bibinfo {volume} {5}},\
  \bibinfo {pages} {2305} (\bibinfo {year} {2018})}\BibitemShut {NoStop}%
\bibitem [{\citenamefont {Chejanovsky}\ \emph {et~al.}(2016)\citenamefont
  {Chejanovsky}, \citenamefont {Rezai}, \citenamefont {Paolucci}, \citenamefont
  {Kim}, \citenamefont {Rendler}, \citenamefont {Rouabeh}, \citenamefont
  {F{\'a}varo~de Oliveira}, \citenamefont {Herlinger}, \citenamefont
  {Denisenko}, \citenamefont {Yang}, \citenamefont {Gerhardt}, \citenamefont
  {Finkler}, \citenamefont {Smet},\ and\ \citenamefont
  {Wrachtrup}}]{10.1021/acs.nanolett.6b03268}%
  \BibitemOpen
  \bibfield  {author} {\bibinfo {author} {\bibfnamefont {N.}~\bibnamefont
  {Chejanovsky}}, \bibinfo {author} {\bibfnamefont {M.}~\bibnamefont {Rezai}},
  \bibinfo {author} {\bibfnamefont {F.}~\bibnamefont {Paolucci}}, \bibinfo
  {author} {\bibfnamefont {Y.}~\bibnamefont {Kim}}, \bibinfo {author}
  {\bibfnamefont {T.}~\bibnamefont {Rendler}}, \bibinfo {author} {\bibfnamefont
  {W.}~\bibnamefont {Rouabeh}}, \bibinfo {author} {\bibfnamefont
  {F.}~\bibnamefont {F{\'a}varo~de Oliveira}}, \bibinfo {author} {\bibfnamefont
  {P.}~\bibnamefont {Herlinger}}, \bibinfo {author} {\bibfnamefont
  {A.}~\bibnamefont {Denisenko}}, \bibinfo {author} {\bibfnamefont
  {S.}~\bibnamefont {Yang}}, \bibinfo {author} {\bibfnamefont {I.}~\bibnamefont
  {Gerhardt}}, \bibinfo {author} {\bibfnamefont {A.}~\bibnamefont {Finkler}},
  \bibinfo {author} {\bibfnamefont {J.~H.}\ \bibnamefont {Smet}},\ and\
  \bibinfo {author} {\bibfnamefont {J.}~\bibnamefont {Wrachtrup}},\ }\bibfield
  {title} {\bibinfo {title} {Structural attributes and photodynamics of visible
  spectrum quantum emitters in hexagonal boron nitride},\ }\href
  {https://doi.org/10.1021/acs.nanolett.6b03268} {\bibfield  {journal}
  {\bibinfo  {journal} {Nano Lett.}\ }\textbf {\bibinfo {volume} {16}},\
  \bibinfo {pages} {7037} (\bibinfo {year} {2016})}\BibitemShut {NoStop}%
\bibitem [{\citenamefont {Vogl}\ \emph
  {et~al.}(2019{\natexlab{b}})\citenamefont {Vogl}, \citenamefont {Sripathy},
  \citenamefont {Sharma}, \citenamefont {Reddy}, \citenamefont {Sullivan},
  \citenamefont {Machacek}, \citenamefont {Zhang}, \citenamefont {Karouta},
  \citenamefont {Buchler}, \citenamefont {Doherty}, \citenamefont {Lu},\ and\
  \citenamefont {Lam}}]{10.1038/s41467-019-09219-5}%
  \BibitemOpen
  \bibfield  {author} {\bibinfo {author} {\bibfnamefont {T.}~\bibnamefont
  {Vogl}}, \bibinfo {author} {\bibfnamefont {K.}~\bibnamefont {Sripathy}},
  \bibinfo {author} {\bibfnamefont {A.}~\bibnamefont {Sharma}}, \bibinfo
  {author} {\bibfnamefont {P.}~\bibnamefont {Reddy}}, \bibinfo {author}
  {\bibfnamefont {J.}~\bibnamefont {Sullivan}}, \bibinfo {author}
  {\bibfnamefont {J.~R.}\ \bibnamefont {Machacek}}, \bibinfo {author}
  {\bibfnamefont {L.}~\bibnamefont {Zhang}}, \bibinfo {author} {\bibfnamefont
  {F.}~\bibnamefont {Karouta}}, \bibinfo {author} {\bibfnamefont {B.~C.}\
  \bibnamefont {Buchler}}, \bibinfo {author} {\bibfnamefont {M.~W.}\
  \bibnamefont {Doherty}}, \bibinfo {author} {\bibfnamefont {Y.}~\bibnamefont
  {Lu}},\ and\ \bibinfo {author} {\bibfnamefont {P.~K.}\ \bibnamefont {Lam}},\
  }\bibfield  {title} {\bibinfo {title} {Radiation tolerance of two-dimensional
  material-based devices for space applications},\ }\href
  {https://doi.org/10.1038/s41467-019-09219-5} {\bibfield  {journal} {\bibinfo
  {journal} {Nat. Commun.}\ }\textbf {\bibinfo {volume} {10}},\ \bibinfo
  {pages} {1202} (\bibinfo {year} {2019}{\natexlab{b}})}\BibitemShut {NoStop}%
\bibitem [{\citenamefont {Proscia}\ \emph {et~al.}(2018)\citenamefont
  {Proscia}, \citenamefont {Shotan}, \citenamefont {Jayakumar}, \citenamefont
  {Reddy}, \citenamefont {Cohen}, \citenamefont {Dollar}, \citenamefont
  {Alkauskas}, \citenamefont {Doherty}, \citenamefont {Meriles},\ and\
  \citenamefont {Menon}}]{10.1364/OPTICA.5.001128}%
  \BibitemOpen
  \bibfield  {author} {\bibinfo {author} {\bibfnamefont {N.~V.}\ \bibnamefont
  {Proscia}}, \bibinfo {author} {\bibfnamefont {Z.}~\bibnamefont {Shotan}},
  \bibinfo {author} {\bibfnamefont {H.}~\bibnamefont {Jayakumar}}, \bibinfo
  {author} {\bibfnamefont {P.}~\bibnamefont {Reddy}}, \bibinfo {author}
  {\bibfnamefont {C.}~\bibnamefont {Cohen}}, \bibinfo {author} {\bibfnamefont
  {M.}~\bibnamefont {Dollar}}, \bibinfo {author} {\bibfnamefont
  {A.}~\bibnamefont {Alkauskas}}, \bibinfo {author} {\bibfnamefont
  {M.}~\bibnamefont {Doherty}}, \bibinfo {author} {\bibfnamefont {C.~A.}\
  \bibnamefont {Meriles}},\ and\ \bibinfo {author} {\bibfnamefont {V.~M.}\
  \bibnamefont {Menon}},\ }\bibfield  {title} {\bibinfo {title}
  {Near-deterministic activation of room-temperature quantum emitters in
  hexagonal boron nitride},\ }\href {https://doi.org/10.1364/OPTICA.5.001128}
  {\bibfield  {journal} {\bibinfo  {journal} {Optica}\ }\textbf {\bibinfo
  {volume} {5}},\ \bibinfo {pages} {1128} (\bibinfo {year} {2018})}\BibitemShut
  {NoStop}%
\bibitem [{\citenamefont {Xu}\ \emph {et~al.}(2021)\citenamefont {Xu},
  \citenamefont {Martin}, \citenamefont {Sychev}, \citenamefont {Lagutchev},
  \citenamefont {Chen}, \citenamefont {Taniguchi}, \citenamefont {Watanabe},
  \citenamefont {Shalaev},\ and\ \citenamefont
  {Boltasseva}}]{doi:10.1021/acs.nanolett.1c02640}%
  \BibitemOpen
  \bibfield  {author} {\bibinfo {author} {\bibfnamefont {X.}~\bibnamefont
  {Xu}}, \bibinfo {author} {\bibfnamefont {Z.~O.}\ \bibnamefont {Martin}},
  \bibinfo {author} {\bibfnamefont {D.}~\bibnamefont {Sychev}}, \bibinfo
  {author} {\bibfnamefont {A.~S.}\ \bibnamefont {Lagutchev}}, \bibinfo {author}
  {\bibfnamefont {Y.~P.}\ \bibnamefont {Chen}}, \bibinfo {author}
  {\bibfnamefont {T.}~\bibnamefont {Taniguchi}}, \bibinfo {author}
  {\bibfnamefont {K.}~\bibnamefont {Watanabe}}, \bibinfo {author}
  {\bibfnamefont {V.~M.}\ \bibnamefont {Shalaev}},\ and\ \bibinfo {author}
  {\bibfnamefont {A.}~\bibnamefont {Boltasseva}},\ }\bibfield  {title}
  {\bibinfo {title} {Creating quantum emitters in hexagonal boron nitride
  deterministically on chip-compatible substrates},\ }\href
  {https://doi.org/10.1021/acs.nanolett.1c02640} {\bibfield  {journal}
  {\bibinfo  {journal} {Nano Lett.}\ }\textbf {\bibinfo {volume} {21}},\
  \bibinfo {pages} {8182} (\bibinfo {year} {2021})}\BibitemShut {NoStop}%
\bibitem [{\citenamefont {Fournier}\ \emph {et~al.}(2021)\citenamefont
  {Fournier}, \citenamefont {Plaud}, \citenamefont {Roux}, \citenamefont
  {Pierret}, \citenamefont {Rosticher}, \citenamefont {Watanabe}, \citenamefont
  {Taniguchi}, \citenamefont {Buil}, \citenamefont {Quélin}, \citenamefont
  {Barjon},\ and\ \citenamefont {et~al.}}]{naturecomm}%
  \BibitemOpen
  \bibfield  {author} {\bibinfo {author} {\bibfnamefont {C.}~\bibnamefont
  {Fournier}}, \bibinfo {author} {\bibfnamefont {A.}~\bibnamefont {Plaud}},
  \bibinfo {author} {\bibfnamefont {S.}~\bibnamefont {Roux}}, \bibinfo {author}
  {\bibfnamefont {A.}~\bibnamefont {Pierret}}, \bibinfo {author} {\bibfnamefont
  {M.}~\bibnamefont {Rosticher}}, \bibinfo {author} {\bibfnamefont
  {K.}~\bibnamefont {Watanabe}}, \bibinfo {author} {\bibfnamefont
  {T.}~\bibnamefont {Taniguchi}}, \bibinfo {author} {\bibfnamefont
  {S.}~\bibnamefont {Buil}}, \bibinfo {author} {\bibfnamefont {X.}~\bibnamefont
  {Quélin}}, \bibinfo {author} {\bibfnamefont {J.}~\bibnamefont {Barjon}},\
  and\ \bibinfo {author} {\bibnamefont {et~al.}},\ }\bibfield  {title}
  {\bibinfo {title} {Position-controlled quantum emitters with reproducible
  emission wavelength in hexagonal boron nitride},\ }\href
  {https://doi.org/10.1038/s41467-021-24019-6} {\bibfield  {journal} {\bibinfo
  {journal} {Nat. Commun.}\ }\textbf {\bibinfo {volume} {12}},\ \bibinfo
  {pages} {3779} (\bibinfo {year} {2021})}\BibitemShut {NoStop}%
\bibitem [{\citenamefont {Gale}\ \emph {et~al.}(2022)\citenamefont {Gale},
  \citenamefont {Li}, \citenamefont {Chen}, \citenamefont {Watanabe},
  \citenamefont {Taniguchi}, \citenamefont {Aharonovich},\ and\ \citenamefont
  {Toth}}]{Gale_2022}%
  \BibitemOpen
  \bibfield  {author} {\bibinfo {author} {\bibfnamefont {A.}~\bibnamefont
  {Gale}}, \bibinfo {author} {\bibfnamefont {C.}~\bibnamefont {Li}}, \bibinfo
  {author} {\bibfnamefont {Y.}~\bibnamefont {Chen}}, \bibinfo {author}
  {\bibfnamefont {K.}~\bibnamefont {Watanabe}}, \bibinfo {author}
  {\bibfnamefont {T.}~\bibnamefont {Taniguchi}}, \bibinfo {author}
  {\bibfnamefont {I.}~\bibnamefont {Aharonovich}},\ and\ \bibinfo {author}
  {\bibfnamefont {M.}~\bibnamefont {Toth}},\ }\bibfield  {title} {\bibinfo
  {title} {Site-specific fabrication of blue quantum emitters in hexagonal
  boron nitride},\ }\href {https://doi.org/10.1021/acsphotonics.2c00631}
  {\bibfield  {journal} {\bibinfo  {journal} {ACS Photonics}\ }\textbf
  {\bibinfo {volume} {9}},\ \bibinfo {pages} {2170–2177} (\bibinfo {year}
  {2022})}\BibitemShut {NoStop}%
\bibitem [{\citenamefont {Kumar}\ \emph {et~al.}(2023)\citenamefont {Kumar},
  \citenamefont {Cholsuk}, \citenamefont {Zand}, \citenamefont {Mishuk},
  \citenamefont {Matthes}, \citenamefont {Eilenberger}, \citenamefont
  {Suwanna},\ and\ \citenamefont {Vogl}}]{akumar}%
  \BibitemOpen
  \bibfield  {author} {\bibinfo {author} {\bibfnamefont {A.}~\bibnamefont
  {Kumar}}, \bibinfo {author} {\bibfnamefont {C.}~\bibnamefont {Cholsuk}},
  \bibinfo {author} {\bibfnamefont {A.}~\bibnamefont {Zand}}, \bibinfo {author}
  {\bibfnamefont {M.~N.}\ \bibnamefont {Mishuk}}, \bibinfo {author}
  {\bibfnamefont {T.}~\bibnamefont {Matthes}}, \bibinfo {author} {\bibfnamefont
  {F.}~\bibnamefont {Eilenberger}}, \bibinfo {author} {\bibfnamefont
  {S.}~\bibnamefont {Suwanna}},\ and\ \bibinfo {author} {\bibfnamefont
  {T.}~\bibnamefont {Vogl}},\ }\bibfield  {title} {\bibinfo {title} {{Localized
  creation of yellow single photon emitting carbon complexes in hexagonal boron
  nitride}},\ }\href {https://doi.org/10.1063/5.0147560} {\bibfield  {journal}
  {\bibinfo  {journal} {APL Mater.}\ }\textbf {\bibinfo {volume} {11}},\
  \bibinfo {pages} {071108} (\bibinfo {year} {2023})}\BibitemShut {NoStop}%
\bibitem [{\citenamefont {Gottscholl}\ \emph {et~al.}(2020)\citenamefont
  {Gottscholl}, \citenamefont {Kianinia}, \citenamefont {Soltamov},
  \citenamefont {Orlinskii}, \citenamefont {Mamin}, \citenamefont {Bradac},
  \citenamefont {Kasper}, \citenamefont {Krambrock}, \citenamefont {Sperlich},
  \citenamefont {Toth}, \citenamefont {Aharonovich},\ and\ \citenamefont
  {Dyakonov}}]{Gottscholl2020}%
  \BibitemOpen
  \bibfield  {author} {\bibinfo {author} {\bibfnamefont {A.}~\bibnamefont
  {Gottscholl}}, \bibinfo {author} {\bibfnamefont {M.}~\bibnamefont
  {Kianinia}}, \bibinfo {author} {\bibfnamefont {V.}~\bibnamefont {Soltamov}},
  \bibinfo {author} {\bibfnamefont {S.}~\bibnamefont {Orlinskii}}, \bibinfo
  {author} {\bibfnamefont {G.}~\bibnamefont {Mamin}}, \bibinfo {author}
  {\bibfnamefont {C.}~\bibnamefont {Bradac}}, \bibinfo {author} {\bibfnamefont
  {C.}~\bibnamefont {Kasper}}, \bibinfo {author} {\bibfnamefont
  {K.}~\bibnamefont {Krambrock}}, \bibinfo {author} {\bibfnamefont
  {A.}~\bibnamefont {Sperlich}}, \bibinfo {author} {\bibfnamefont
  {M.}~\bibnamefont {Toth}}, \bibinfo {author} {\bibfnamefont {I.}~\bibnamefont
  {Aharonovich}},\ and\ \bibinfo {author} {\bibfnamefont {V.}~\bibnamefont
  {Dyakonov}},\ }\bibfield  {title} {\bibinfo {title} {Initialization and
  read-out of intrinsic spin defects in a van der waals crystal at room
  temperature},\ }\href {https://doi.org/10.1038/s41563-020-0619-6} {\bibfield
  {journal} {\bibinfo  {journal} {Nat. Mater.}\ }\textbf {\bibinfo {volume}
  {19}},\ \bibinfo {pages} {540} (\bibinfo {year} {2020})}\BibitemShut
  {NoStop}%
\bibitem [{\citenamefont {Mendelson}\ \emph {et~al.}(2021)\citenamefont
  {Mendelson}, \citenamefont {Chugh}, \citenamefont {Reimers}, \citenamefont
  {Cheng}, \citenamefont {Gottscholl}, \citenamefont {Long}, \citenamefont
  {Mellor}, \citenamefont {Zettl}, \citenamefont {Dyakonov}, \citenamefont
  {Beton}, \citenamefont {Novikov}, \citenamefont {Jagadish}, \citenamefont
  {Tan}, \citenamefont {Ford}, \citenamefont {Toth}, \citenamefont {Bradac},\
  and\ \citenamefont {Aharonovich}}]{Mendelson2021}%
  \BibitemOpen
  \bibfield  {author} {\bibinfo {author} {\bibfnamefont {N.}~\bibnamefont
  {Mendelson}}, \bibinfo {author} {\bibfnamefont {D.}~\bibnamefont {Chugh}},
  \bibinfo {author} {\bibfnamefont {J.~R.}\ \bibnamefont {Reimers}}, \bibinfo
  {author} {\bibfnamefont {T.~S.}\ \bibnamefont {Cheng}}, \bibinfo {author}
  {\bibfnamefont {A.}~\bibnamefont {Gottscholl}}, \bibinfo {author}
  {\bibfnamefont {H.}~\bibnamefont {Long}}, \bibinfo {author} {\bibfnamefont
  {C.~J.}\ \bibnamefont {Mellor}}, \bibinfo {author} {\bibfnamefont
  {A.}~\bibnamefont {Zettl}}, \bibinfo {author} {\bibfnamefont
  {V.}~\bibnamefont {Dyakonov}}, \bibinfo {author} {\bibfnamefont {P.~H.}\
  \bibnamefont {Beton}}, \bibinfo {author} {\bibfnamefont {S.~V.}\ \bibnamefont
  {Novikov}}, \bibinfo {author} {\bibfnamefont {C.}~\bibnamefont {Jagadish}},
  \bibinfo {author} {\bibfnamefont {H.~H.}\ \bibnamefont {Tan}}, \bibinfo
  {author} {\bibfnamefont {M.~J.}\ \bibnamefont {Ford}}, \bibinfo {author}
  {\bibfnamefont {M.}~\bibnamefont {Toth}}, \bibinfo {author} {\bibfnamefont
  {C.}~\bibnamefont {Bradac}},\ and\ \bibinfo {author} {\bibfnamefont
  {I.}~\bibnamefont {Aharonovich}},\ }\bibfield  {title} {\bibinfo {title}
  {Identifying carbon as the source of visible single-photon emission from
  hexagonal boron nitride},\ }\href
  {https://doi.org/10.1038/s41563-020-00850-y} {\bibfield  {journal} {\bibinfo
  {journal} {Nat. Mater.}\ }\textbf {\bibinfo {volume} {20}},\ \bibinfo {pages}
  {321} (\bibinfo {year} {2021})}\BibitemShut {NoStop}%
\bibitem [{\citenamefont {Jungwirth}\ \emph {et~al.}(2016)\citenamefont
  {Jungwirth}, \citenamefont {Calderon}, \citenamefont {Ji}, \citenamefont
  {Spencer}, \citenamefont {Flatté},\ and\ \citenamefont {Fuchs}}]{Jungwirth}%
  \BibitemOpen
  \bibfield  {author} {\bibinfo {author} {\bibfnamefont {N.~R.}\ \bibnamefont
  {Jungwirth}}, \bibinfo {author} {\bibfnamefont {B.}~\bibnamefont {Calderon}},
  \bibinfo {author} {\bibfnamefont {Y.}~\bibnamefont {Ji}}, \bibinfo {author}
  {\bibfnamefont {M.~G.}\ \bibnamefont {Spencer}}, \bibinfo {author}
  {\bibfnamefont {M.~E.}\ \bibnamefont {Flatté}},\ and\ \bibinfo {author}
  {\bibfnamefont {G.~D.}\ \bibnamefont {Fuchs}},\ }\bibfield  {title} {\bibinfo
  {title} {Temperature dependence of wavelength selectable zero-phonon emission
  from single defects in hexagonal boron nitride},\ }\href
  {https://doi.org/10.1021/acs.nanolett.6b01987} {\bibfield  {journal}
  {\bibinfo  {journal} {Nano Letters}\ }\textbf {\bibinfo {volume} {16}},\
  \bibinfo {pages} {6052} (\bibinfo {year} {2016})},\ \bibinfo {note} {pMID:
  27580074},\ \Eprint
  {https://arxiv.org/abs/https://doi.org/10.1021/acs.nanolett.6b01987}
  {https://doi.org/10.1021/acs.nanolett.6b01987} \BibitemShut {NoStop}%
\bibitem [{\citenamefont {Horder}\ \emph {et~al.}(2022)\citenamefont {Horder},
  \citenamefont {White}, \citenamefont {Gale}, \citenamefont {Li},
  \citenamefont {Watanabe}, \citenamefont {Taniguchi}, \citenamefont
  {Kianinia}, \citenamefont {Aharonovich},\ and\ \citenamefont
  {Toth}}]{PhysRevApplied.18.064021}%
  \BibitemOpen
  \bibfield  {author} {\bibinfo {author} {\bibfnamefont {J.}~\bibnamefont
  {Horder}}, \bibinfo {author} {\bibfnamefont {S.~J.}\ \bibnamefont {White}},
  \bibinfo {author} {\bibfnamefont {A.}~\bibnamefont {Gale}}, \bibinfo {author}
  {\bibfnamefont {C.}~\bibnamefont {Li}}, \bibinfo {author} {\bibfnamefont
  {K.}~\bibnamefont {Watanabe}}, \bibinfo {author} {\bibfnamefont
  {T.}~\bibnamefont {Taniguchi}}, \bibinfo {author} {\bibfnamefont
  {M.}~\bibnamefont {Kianinia}}, \bibinfo {author} {\bibfnamefont
  {I.}~\bibnamefont {Aharonovich}},\ and\ \bibinfo {author} {\bibfnamefont
  {M.}~\bibnamefont {Toth}},\ }\bibfield  {title} {\bibinfo {title} {Coherence
  properties of electron-beam-activated emitters in hexagonal boron nitride
  under resonant excitation},\ }\href
  {https://doi.org/10.1103/PhysRevApplied.18.064021} {\bibfield  {journal}
  {\bibinfo  {journal} {Phys. Rev. Appl.}\ }\textbf {\bibinfo {volume} {18}},\
  \bibinfo {pages} {064021} (\bibinfo {year} {2022})}\BibitemShut {NoStop}%
\bibitem [{\citenamefont {Yim}\ \emph {et~al.}(2020)\citenamefont {Yim},
  \citenamefont {Yu}, \citenamefont {Noh}, \citenamefont {Lee},\ and\
  \citenamefont {Seo}}]{yim_et_al}%
  \BibitemOpen
  \bibfield  {author} {\bibinfo {author} {\bibfnamefont {D.}~\bibnamefont
  {Yim}}, \bibinfo {author} {\bibfnamefont {M.}~\bibnamefont {Yu}}, \bibinfo
  {author} {\bibfnamefont {G.}~\bibnamefont {Noh}}, \bibinfo {author}
  {\bibfnamefont {J.}~\bibnamefont {Lee}},\ and\ \bibinfo {author}
  {\bibfnamefont {H.}~\bibnamefont {Seo}},\ }\bibfield  {title} {\bibinfo
  {title} {Polarization and localization of single-photon emitters in hexagonal
  boron nitride wrinkles},\ }\href {https://doi.org/10.1021/acsami.0c09740}
  {\bibfield  {journal} {\bibinfo  {journal} {ACS Appl. Mater. Interfaces.}\
  }\textbf {\bibinfo {volume} {12}},\ \bibinfo {pages} {36362} (\bibinfo {year}
  {2020})},\ \bibinfo {note} {pMID: 32677428}\BibitemShut {NoStop}%
\bibitem [{\citenamefont {Mendelson}\ \emph {et~al.}(2020)\citenamefont
  {Mendelson}, \citenamefont {Doherty}, \citenamefont {Toth}, \citenamefont
  {Aharonovich},\ and\ \citenamefont {Tran}}]{Mendelson2020}%
  \BibitemOpen
  \bibfield  {author} {\bibinfo {author} {\bibfnamefont {N.}~\bibnamefont
  {Mendelson}}, \bibinfo {author} {\bibfnamefont {M.}~\bibnamefont {Doherty}},
  \bibinfo {author} {\bibfnamefont {M.}~\bibnamefont {Toth}}, \bibinfo {author}
  {\bibfnamefont {I.}~\bibnamefont {Aharonovich}},\ and\ \bibinfo {author}
  {\bibfnamefont {T.~T.}\ \bibnamefont {Tran}},\ }\bibfield  {title} {\bibinfo
  {title} {Strain‐induced modification of the optical characteristics of
  quantum emitters in hexagonal boron nitride},\ }\href
  {https://doi.org/10.1002/adma.201908316} {\bibfield  {journal} {\bibinfo
  {journal} {Adv. Mater.}\ }\textbf {\bibinfo {volume} {32}},\ \bibinfo {pages}
  {1908316} (\bibinfo {year} {2020})}\BibitemShut {NoStop}%
\bibitem [{\citenamefont {Iv{\'a}dy}\ \emph {et~al.}(2020)\citenamefont
  {Iv{\'a}dy}, \citenamefont {Barcza}, \citenamefont {Thiering}, \citenamefont
  {Li}, \citenamefont {Hamdi}, \citenamefont {Chou}, \citenamefont {Legeza},\
  and\ \citenamefont {Gali}}]{Ivady2020}%
  \BibitemOpen
  \bibfield  {author} {\bibinfo {author} {\bibfnamefont {V.}~\bibnamefont
  {Iv{\'a}dy}}, \bibinfo {author} {\bibfnamefont {G.}~\bibnamefont {Barcza}},
  \bibinfo {author} {\bibfnamefont {G.}~\bibnamefont {Thiering}}, \bibinfo
  {author} {\bibfnamefont {S.}~\bibnamefont {Li}}, \bibinfo {author}
  {\bibfnamefont {H.}~\bibnamefont {Hamdi}}, \bibinfo {author} {\bibfnamefont
  {J.-P.}\ \bibnamefont {Chou}}, \bibinfo {author} {\bibfnamefont
  {{\"O}.}~\bibnamefont {Legeza}},\ and\ \bibinfo {author} {\bibfnamefont
  {A.}~\bibnamefont {Gali}},\ }\bibfield  {title} {\bibinfo {title} {Ab initio
  theory of the negatively charged boron vacancy qubit in hexagonal boron
  nitride},\ }\href {https://doi.org/10.1038/s41524-020-0305-x} {\bibfield
  {journal} {\bibinfo  {journal} {NPJ Comput. Mater.}\ }\textbf {\bibinfo
  {volume} {6}},\ \bibinfo {pages} {41} (\bibinfo {year} {2020})}\BibitemShut
  {NoStop}%
\bibitem [{\citenamefont {Tawfik}\ \emph {et~al.}(2017)\citenamefont {Tawfik},
  \citenamefont {Ali}, \citenamefont {Fronzi}, \citenamefont {Kianinia},
  \citenamefont {Tran}, \citenamefont {Stampfl}, \citenamefont {Aharonovich},
  \citenamefont {Toth},\ and\ \citenamefont
  {Ford}}]{tawfik_ali_fronzi_kianinia_tran_stampfl_aharonovich_toth_ford_2017}%
  \BibitemOpen
  \bibfield  {author} {\bibinfo {author} {\bibfnamefont {S.~A.}\ \bibnamefont
  {Tawfik}}, \bibinfo {author} {\bibfnamefont {S.}~\bibnamefont {Ali}},
  \bibinfo {author} {\bibfnamefont {M.}~\bibnamefont {Fronzi}}, \bibinfo
  {author} {\bibfnamefont {M.}~\bibnamefont {Kianinia}}, \bibinfo {author}
  {\bibfnamefont {T.~T.}\ \bibnamefont {Tran}}, \bibinfo {author}
  {\bibfnamefont {C.}~\bibnamefont {Stampfl}}, \bibinfo {author} {\bibfnamefont
  {I.}~\bibnamefont {Aharonovich}}, \bibinfo {author} {\bibfnamefont
  {M.}~\bibnamefont {Toth}},\ and\ \bibinfo {author} {\bibfnamefont {M.~J.}\
  \bibnamefont {Ford}},\ }\bibfield  {title} {\bibinfo {title}
  {First-principles investigation of quantum emission from hbn defects},\
  }\href {https://doi.org/10.1039/c7nr04270a} {\bibfield  {journal} {\bibinfo
  {journal} {Nanoscale}\ }\textbf {\bibinfo {volume} {9}},\ \bibinfo {pages}
  {13575–13582} (\bibinfo {year} {2017})}\BibitemShut {NoStop}%
\bibitem [{\citenamefont {Ates}\ \emph {et~al.}(2013)\citenamefont {Ates},
  \citenamefont {Agha}, \citenamefont {Gulinatti}, \citenamefont {Rech},
  \citenamefont {Badolato},\ and\ \citenamefont {Srinivasan}}]{Ates2013}%
  \BibitemOpen
  \bibfield  {author} {\bibinfo {author} {\bibfnamefont {S.}~\bibnamefont
  {Ates}}, \bibinfo {author} {\bibfnamefont {I.}~\bibnamefont {Agha}}, \bibinfo
  {author} {\bibfnamefont {A.}~\bibnamefont {Gulinatti}}, \bibinfo {author}
  {\bibfnamefont {I.}~\bibnamefont {Rech}}, \bibinfo {author} {\bibfnamefont
  {A.}~\bibnamefont {Badolato}},\ and\ \bibinfo {author} {\bibfnamefont
  {K.}~\bibnamefont {Srinivasan}},\ }\bibfield  {title} {\bibinfo {title}
  {Improving the performance of bright quantum dot single photon sources using
  temporal filtering via amplitude modulation},\ }\href
  {https://doi.org/10.1038/srep01397} {\bibfield  {journal} {\bibinfo
  {journal} {Sci. Rep.}\ }\textbf {\bibinfo {volume} {3}},\ \bibinfo {pages}
  {1} (\bibinfo {year} {2013})}\BibitemShut {NoStop}%
\bibitem [{\citenamefont {Samaner}\ \emph {et~al.}(2022)\citenamefont
  {Samaner}, \citenamefont {Pacal}, \citenamefont {Mutlu}, \citenamefont
  {Uyanik},\ and\ \citenamefont {Ates}}]{Caglar}%
  \BibitemOpen
  \bibfield  {author} {\bibinfo {author} {\bibfnamefont {C.}~\bibnamefont
  {Samaner}}, \bibinfo {author} {\bibfnamefont {S.}~\bibnamefont {Pacal}},
  \bibinfo {author} {\bibfnamefont {G.}~\bibnamefont {Mutlu}}, \bibinfo
  {author} {\bibfnamefont {K.}~\bibnamefont {Uyanik}},\ and\ \bibinfo {author}
  {\bibfnamefont {S.}~\bibnamefont {Ates}},\ }\bibfield  {title} {\bibinfo
  {title} {Free‐space quantum key distribution with single photons from
  defects in hexagonal boron nitride},\ }\href
  {https://doi.org/10.1002/qute.202200059} {\bibfield  {journal} {\bibinfo
  {journal} {Adv. Quantum Technol.}\ }\textbf {\bibinfo {volume} {5}},\
  \bibinfo {pages} {2200059} (\bibinfo {year} {2022})}\BibitemShut {NoStop}%
\bibitem [{\citenamefont {Patel}\ \emph {et~al.}(2022)\citenamefont {Patel},
  \citenamefont {Hopper}, \citenamefont {Gusdorff}, \citenamefont {Turiansky},
  \citenamefont {Huang}, \citenamefont {Fishman}, \citenamefont {Porat},
  \citenamefont {Walle},\ and\ \citenamefont {Bassett}}]{Patel2022}%
  \BibitemOpen
  \bibfield  {author} {\bibinfo {author} {\bibfnamefont {R.~N.}\ \bibnamefont
  {Patel}}, \bibinfo {author} {\bibfnamefont {D.~A.}\ \bibnamefont {Hopper}},
  \bibinfo {author} {\bibfnamefont {J.~A.}\ \bibnamefont {Gusdorff}}, \bibinfo
  {author} {\bibfnamefont {M.~E.}\ \bibnamefont {Turiansky}}, \bibinfo {author}
  {\bibfnamefont {T.~Y.}\ \bibnamefont {Huang}}, \bibinfo {author}
  {\bibfnamefont {R.~E.}\ \bibnamefont {Fishman}}, \bibinfo {author}
  {\bibfnamefont {B.}~\bibnamefont {Porat}}, \bibinfo {author} {\bibfnamefont
  {C.~G. V.~D.}\ \bibnamefont {Walle}},\ and\ \bibinfo {author} {\bibfnamefont
  {L.~C.}\ \bibnamefont {Bassett}},\ }\bibfield  {title} {\bibinfo {title}
  {Probing the optical dynamics of quantum emitters in hexagonal boron
  nitride},\ }\href
  {https://doi.org/https://doi.org/10.1103/PRXQuantum.3.030331} {\bibfield
  {journal} {\bibinfo  {journal} {PRX Quantum}\ }\textbf {\bibinfo {volume}
  {3}},\ \bibinfo {pages} {030331} (\bibinfo {year} {2022})}\BibitemShut
  {NoStop}%
\bibitem [{\citenamefont {Constantinescu}\ \emph {et~al.}(2013)\citenamefont
  {Constantinescu}, \citenamefont {Kuc},\ and\ \citenamefont
  {Heine}}]{Constantinescu2013}%
  \BibitemOpen
  \bibfield  {author} {\bibinfo {author} {\bibfnamefont {G.}~\bibnamefont
  {Constantinescu}}, \bibinfo {author} {\bibfnamefont {A.}~\bibnamefont
  {Kuc}},\ and\ \bibinfo {author} {\bibfnamefont {T.}~\bibnamefont {Heine}},\
  }\bibfield  {title} {\bibinfo {title} {Stacking in bulk and bilayer hexagonal
  boron nitride},\ }\href {https://doi.org/10.1103/PhysRevLett.111.036104}
  {\bibfield  {journal} {\bibinfo  {journal} {Phys. Rev. Lett.}\ }\textbf
  {\bibinfo {volume} {111}},\ \bibinfo {pages} {036104} (\bibinfo {year}
  {2013})}\BibitemShut {NoStop}%
\bibitem [{\citenamefont {Li}\ \emph {et~al.}(2013)\citenamefont {Li},
  \citenamefont {Rao}, \citenamefont {Mak}, \citenamefont {You}, \citenamefont
  {Wang}, \citenamefont {Dean},\ and\ \citenamefont
  {Heinz}}]{doi:10.1021/nl401561r}%
  \BibitemOpen
  \bibfield  {author} {\bibinfo {author} {\bibfnamefont {Y.}~\bibnamefont
  {Li}}, \bibinfo {author} {\bibfnamefont {Y.}~\bibnamefont {Rao}}, \bibinfo
  {author} {\bibfnamefont {K.~F.}\ \bibnamefont {Mak}}, \bibinfo {author}
  {\bibfnamefont {Y.}~\bibnamefont {You}}, \bibinfo {author} {\bibfnamefont
  {S.}~\bibnamefont {Wang}}, \bibinfo {author} {\bibfnamefont {C.~R.}\
  \bibnamefont {Dean}},\ and\ \bibinfo {author} {\bibfnamefont {T.~F.}\
  \bibnamefont {Heinz}},\ }\bibfield  {title} {\bibinfo {title} {Probing
  symmetry properties of few-layer mos2 and h-bn by optical second-harmonic
  generation},\ }\href {https://doi.org/10.1021/nl401561r} {\bibfield
  {journal} {\bibinfo  {journal} {Nano Lett.}\ }\textbf {\bibinfo {volume}
  {13}},\ \bibinfo {pages} {3329} (\bibinfo {year} {2013})}\BibitemShut
  {NoStop}%
\bibitem [{\citenamefont {Mennel}\ \emph {et~al.}(2018)\citenamefont {Mennel},
  \citenamefont {Furchi}, \citenamefont {Wachter}, \citenamefont {Paur},
  \citenamefont {Polyushkin},\ and\ \citenamefont {Mueller}}]{Mennel2018}%
  \BibitemOpen
  \bibfield  {author} {\bibinfo {author} {\bibfnamefont {L.}~\bibnamefont
  {Mennel}}, \bibinfo {author} {\bibfnamefont {M.~M.}\ \bibnamefont {Furchi}},
  \bibinfo {author} {\bibfnamefont {S.}~\bibnamefont {Wachter}}, \bibinfo
  {author} {\bibfnamefont {M.}~\bibnamefont {Paur}}, \bibinfo {author}
  {\bibfnamefont {D.~K.}\ \bibnamefont {Polyushkin}},\ and\ \bibinfo {author}
  {\bibfnamefont {T.}~\bibnamefont {Mueller}},\ }\bibfield  {title} {\bibinfo
  {title} {Optical imaging of strain in two-dimensional crystals},\ }\href
  {https://doi.org/10.1038/s41467-018-02830-y} {\bibfield  {journal} {\bibinfo
  {journal} {Nat. Commun.}\ }\textbf {\bibinfo {volume} {9}},\ \bibinfo {pages}
  {516} (\bibinfo {year} {2018})}\BibitemShut {NoStop}%
\bibitem [{\citenamefont {Weston}\ \emph {et~al.}(2018)\citenamefont {Weston},
  \citenamefont {Wickramaratne}, \citenamefont {Mackoit}, \citenamefont
  {Alkauskas},\ and\ \citenamefont {Van~de Walle}}]{Weston2018}%
  \BibitemOpen
  \bibfield  {author} {\bibinfo {author} {\bibfnamefont {L.}~\bibnamefont
  {Weston}}, \bibinfo {author} {\bibfnamefont {D.}~\bibnamefont
  {Wickramaratne}}, \bibinfo {author} {\bibfnamefont {M.}~\bibnamefont
  {Mackoit}}, \bibinfo {author} {\bibfnamefont {A.}~\bibnamefont {Alkauskas}},\
  and\ \bibinfo {author} {\bibfnamefont {C.~G.}\ \bibnamefont {Van~de Walle}},\
  }\bibfield  {title} {\bibinfo {title} {Native point defects and impurities in
  hexagonal boron nitride},\ }\href
  {https://doi.org/10.1103/PhysRevB.97.214104} {\bibfield  {journal} {\bibinfo
  {journal} {Phys. Rev. B}\ }\textbf {\bibinfo {volume} {97}},\ \bibinfo
  {pages} {214104} (\bibinfo {year} {2018})}\BibitemShut {NoStop}%
\bibitem [{\citenamefont {Ding}\ \emph {et~al.}(2016)\citenamefont {Ding},
  \citenamefont {He}, \citenamefont {Duan}, \citenamefont {Gregersen},
  \citenamefont {Chen}, \citenamefont {Unsleber}, \citenamefont {Maier},
  \citenamefont {Schneider}, \citenamefont {Kamp}, \citenamefont {H\"ofling},
  \citenamefont {Lu},\ and\ \citenamefont {Pan}}]{PhysRevLett.116.020401}%
  \BibitemOpen
  \bibfield  {author} {\bibinfo {author} {\bibfnamefont {X.}~\bibnamefont
  {Ding}}, \bibinfo {author} {\bibfnamefont {Y.}~\bibnamefont {He}}, \bibinfo
  {author} {\bibfnamefont {Z.-C.}\ \bibnamefont {Duan}}, \bibinfo {author}
  {\bibfnamefont {N.}~\bibnamefont {Gregersen}}, \bibinfo {author}
  {\bibfnamefont {M.-C.}\ \bibnamefont {Chen}}, \bibinfo {author}
  {\bibfnamefont {S.}~\bibnamefont {Unsleber}}, \bibinfo {author}
  {\bibfnamefont {S.}~\bibnamefont {Maier}}, \bibinfo {author} {\bibfnamefont
  {C.}~\bibnamefont {Schneider}}, \bibinfo {author} {\bibfnamefont
  {M.}~\bibnamefont {Kamp}}, \bibinfo {author} {\bibfnamefont {S.}~\bibnamefont
  {H\"ofling}}, \bibinfo {author} {\bibfnamefont {C.-Y.}\ \bibnamefont {Lu}},\
  and\ \bibinfo {author} {\bibfnamefont {J.-W.}\ \bibnamefont {Pan}},\
  }\bibfield  {title} {\bibinfo {title} {On-demand single photons with high
  extraction efficiency and near-unity indistinguishability from a resonantly
  driven quantum dot in a micropillar},\ }\href
  {https://doi.org/10.1103/PhysRevLett.116.020401} {\bibfield  {journal}
  {\bibinfo  {journal} {Phys. Rev. Lett.}\ }\textbf {\bibinfo {volume} {116}},\
  \bibinfo {pages} {020401} (\bibinfo {year} {2016})}\BibitemShut {NoStop}%
\bibitem [{\citenamefont {Tonndorf}\ \emph {et~al.}(2015)\citenamefont
  {Tonndorf}, \citenamefont {Schmidt}, \citenamefont {Schneider}, \citenamefont
  {Kern}, \citenamefont {Buscema}, \citenamefont {Steele}, \citenamefont
  {Castellanos-Gomez}, \citenamefont {van~der Zant}, \citenamefont
  {de~Vasconcellos},\ and\ \citenamefont {Bratschitsch}}]{Tonndorf:15}%
  \BibitemOpen
  \bibfield  {author} {\bibinfo {author} {\bibfnamefont {P.}~\bibnamefont
  {Tonndorf}}, \bibinfo {author} {\bibfnamefont {R.}~\bibnamefont {Schmidt}},
  \bibinfo {author} {\bibfnamefont {R.}~\bibnamefont {Schneider}}, \bibinfo
  {author} {\bibfnamefont {J.}~\bibnamefont {Kern}}, \bibinfo {author}
  {\bibfnamefont {M.}~\bibnamefont {Buscema}}, \bibinfo {author} {\bibfnamefont
  {G.~A.}\ \bibnamefont {Steele}}, \bibinfo {author} {\bibfnamefont
  {A.}~\bibnamefont {Castellanos-Gomez}}, \bibinfo {author} {\bibfnamefont
  {H.~S.~J.}\ \bibnamefont {van~der Zant}}, \bibinfo {author} {\bibfnamefont
  {S.~M.}\ \bibnamefont {de~Vasconcellos}},\ and\ \bibinfo {author}
  {\bibfnamefont {R.}~\bibnamefont {Bratschitsch}},\ }\bibfield  {title}
  {\bibinfo {title} {Single-photon emission from localized excitons in an
  atomically thin semiconductor},\ }\href
  {https://doi.org/10.1364/OPTICA.2.000347} {\bibfield  {journal} {\bibinfo
  {journal} {Optica}\ }\textbf {\bibinfo {volume} {2}},\ \bibinfo {pages} {347}
  (\bibinfo {year} {2015})}\BibitemShut {NoStop}%
\bibitem [{\citenamefont {Reimers}\ \emph {et~al.}(2018)\citenamefont
  {Reimers}, \citenamefont {Sajid}, \citenamefont {Kobayashi},\ and\
  \citenamefont {Ford}}]{Reimers2018}%
  \BibitemOpen
  \bibfield  {author} {\bibinfo {author} {\bibfnamefont {J.~R.}\ \bibnamefont
  {Reimers}}, \bibinfo {author} {\bibfnamefont {A.}~\bibnamefont {Sajid}},
  \bibinfo {author} {\bibfnamefont {R.}~\bibnamefont {Kobayashi}},\ and\
  \bibinfo {author} {\bibfnamefont {M.~J.}\ \bibnamefont {Ford}},\ }\bibfield
  {title} {\bibinfo {title} {Understanding and calibrating
  density-functional-theory calculations describing the energy and spectroscopy
  of defect sites in hexagonal boron nitride},\ }\href
  {https://doi.org/10.1021/acs.jctc.7b01072} {\bibfield  {journal} {\bibinfo
  {journal} {J. Chem. Theory Comput.}\ }\textbf {\bibinfo {volume} {14}},\
  \bibinfo {pages} {1602} (\bibinfo {year} {2018})}\BibitemShut {NoStop}%
\bibitem [{\citenamefont {Cholsuk}\ \emph {et~al.}(2022)\citenamefont
  {Cholsuk}, \citenamefont {Suwanna},\ and\ \citenamefont
  {Vogl}}]{Cholsuk2022}%
  \BibitemOpen
  \bibfield  {author} {\bibinfo {author} {\bibfnamefont {C.}~\bibnamefont
  {Cholsuk}}, \bibinfo {author} {\bibfnamefont {S.}~\bibnamefont {Suwanna}},\
  and\ \bibinfo {author} {\bibfnamefont {T.}~\bibnamefont {Vogl}},\ }\bibfield
  {title} {\bibinfo {title} {Tailoring the emission wavelength of color centers
  in hexagonal boron nitride for quantum applications},\ }\href
  {https://doi.org/10.3390/nano12142427} {\bibfield  {journal} {\bibinfo
  {journal} {Nanomaterials}\ }\textbf {\bibinfo {volume} {12}},\ \bibinfo
  {pages} {2427} (\bibinfo {year} {2022})}\BibitemShut {NoStop}%
\bibitem [{\citenamefont {Auburger}\ and\ \citenamefont
  {Gali}(2021)}]{Auburger2021}%
  \BibitemOpen
  \bibfield  {author} {\bibinfo {author} {\bibfnamefont {P.}~\bibnamefont
  {Auburger}}\ and\ \bibinfo {author} {\bibfnamefont {A.}~\bibnamefont
  {Gali}},\ }\bibfield  {title} {\bibinfo {title} {Towards ab initio
  identification of paramagnetic substitutional carbon defects in hexagonal
  boron nitride acting as quantum bits},\ }\href
  {https://doi.org/10.1103/PhysRevB.104.075410} {\bibfield  {journal} {\bibinfo
   {journal} {Phys. Rev. B}\ }\textbf {\bibinfo {volume} {104}},\ \bibinfo
  {pages} {075410} (\bibinfo {year} {2021})}\BibitemShut {NoStop}%
\bibitem [{\citenamefont {Jungwirth}\ and\ \citenamefont
  {Fuchs}(2017)}]{Jungwirth2017}%
  \BibitemOpen
  \bibfield  {author} {\bibinfo {author} {\bibfnamefont {N.~R.}\ \bibnamefont
  {Jungwirth}}\ and\ \bibinfo {author} {\bibfnamefont {G.~D.}\ \bibnamefont
  {Fuchs}},\ }\bibfield  {title} {\bibinfo {title} {Optical absorption and
  emission mechanisms of single defects in hexagonal boron nitride},\ }\href
  {https://doi.org/10.1103/PhysRevLett.119.057401} {\bibfield  {journal}
  {\bibinfo  {journal} {Phys. Rev. Lett.}\ }\textbf {\bibinfo {volume} {119}},\
  \bibinfo {pages} {057401} (\bibinfo {year} {2017})}\BibitemShut {NoStop}%
\bibitem [{\citenamefont {Davidsson}(2020)}]{Davidsson2020}%
  \BibitemOpen
  \bibfield  {author} {\bibinfo {author} {\bibfnamefont {J.}~\bibnamefont
  {Davidsson}},\ }\bibfield  {title} {\bibinfo {title} {Theoretical
  polarization of zero phonon lines in point defects},\ }\href
  {https://doi.org/10.1088/1361-648X/ab94f4} {\bibfield  {journal} {\bibinfo
  {journal} {J. Phys.: Condens. Matter}\ }\textbf {\bibinfo {volume} {32}},\
  \bibinfo {pages} {385502} (\bibinfo {year} {2020})}\BibitemShut {NoStop}%
\bibitem [{\citenamefont {Li}\ \emph {et~al.}(2022)\citenamefont {Li},
  \citenamefont {Smart},\ and\ \citenamefont {Ping}}]{Li2022}%
  \BibitemOpen
  \bibfield  {author} {\bibinfo {author} {\bibfnamefont {K.}~\bibnamefont
  {Li}}, \bibinfo {author} {\bibfnamefont {T.~J.}\ \bibnamefont {Smart}},\ and\
  \bibinfo {author} {\bibfnamefont {Y.}~\bibnamefont {Ping}},\ }\bibfield
  {title} {\bibinfo {title} {Carbon trimer as a 2 ev single-photon emitter
  candidate in hexagonal boron nitride: A first-principles study},\ }\href
  {https://doi.org/10.1103/PhysRevMaterials.6.L042201} {\bibfield  {journal}
  {\bibinfo  {journal} {Phys. Rev. Mater.}\ }\textbf {\bibinfo {volume} {6}},\
  \bibinfo {pages} {L042201} (\bibinfo {year} {2022})}\BibitemShut {NoStop}%
\bibitem [{\citenamefont {Sajid}\ and\ \citenamefont
  {Thygesen}(2020)}]{Sajid2020-2}%
  \BibitemOpen
  \bibfield  {author} {\bibinfo {author} {\bibfnamefont {A.}~\bibnamefont
  {Sajid}}\ and\ \bibinfo {author} {\bibfnamefont {K.~S.}\ \bibnamefont
  {Thygesen}},\ }\bibfield  {title} {\bibinfo {title}
  {V$_\text{N}$\text{C}$_\text{B}$ defect as source of single photon emission
  from hexagonal boron nitride},\ }\href
  {https://doi.org/10.1088/2053-1583/ab8f61} {\bibfield  {journal} {\bibinfo
  {journal} {2D Mater.}\ }\textbf {\bibinfo {volume} {7}},\ \bibinfo {pages}
  {031007} (\bibinfo {year} {2020})}\BibitemShut {NoStop}%
\bibitem [{\citenamefont {Cholsuk}\ \emph {et~al.}(2023)\citenamefont
  {Cholsuk}, \citenamefont {Suwanna},\ and\ \citenamefont
  {Vogl}}]{cholsuk2023-comprehensive}%
  \BibitemOpen
  \bibfield  {author} {\bibinfo {author} {\bibfnamefont {C.}~\bibnamefont
  {Cholsuk}}, \bibinfo {author} {\bibfnamefont {S.}~\bibnamefont {Suwanna}},\
  and\ \bibinfo {author} {\bibfnamefont {T.}~\bibnamefont {Vogl}},\ }\bibfield
  {title} {\bibinfo {title} {Comprehensive scheme for identifying defects in
  solid-state quantum systems},\ }\href
  {https://doi.org/10.1021/acs.jpclett.3c01475} {\bibfield  {journal} {\bibinfo
   {journal} {The Journal of Physical Chemistry Letters}\ }\textbf {\bibinfo
  {volume} {14}},\ \bibinfo {pages} {6564} (\bibinfo {year}
  {2023})}\BibitemShut {NoStop}%
\bibitem [{\citenamefont {Srivastava}\ \emph {et~al.}(2015)\citenamefont
  {Srivastava}, \citenamefont {Sidler}, \citenamefont {Allain}, \citenamefont
  {Lembke}, \citenamefont {Kis},\ and\ \citenamefont
  {Imamo\v{g}lu}}]{nnano.2015.60}%
  \BibitemOpen
  \bibfield  {author} {\bibinfo {author} {\bibfnamefont {A.}~\bibnamefont
  {Srivastava}}, \bibinfo {author} {\bibfnamefont {M.}~\bibnamefont {Sidler}},
  \bibinfo {author} {\bibfnamefont {A.~V.}\ \bibnamefont {Allain}}, \bibinfo
  {author} {\bibfnamefont {D.~S.}\ \bibnamefont {Lembke}}, \bibinfo {author}
  {\bibfnamefont {A.}~\bibnamefont {Kis}},\ and\ \bibinfo {author}
  {\bibfnamefont {A.}~\bibnamefont {Imamo\v{g}lu}},\ }\bibfield  {title}
  {\bibinfo {title} {Optically active quantum dots in monolayer wse2},\ }\href
  {https://doi.org/10.1038/nnano.2015.60} {\bibfield  {journal} {\bibinfo
  {journal} {Nat. Nanotechnol.}\ }\textbf {\bibinfo {volume} {10}},\ \bibinfo
  {pages} {491} (\bibinfo {year} {2015})}\BibitemShut {NoStop}%
\bibitem [{\citenamefont {He}\ \emph {et~al.}(2015)\citenamefont {He},
  \citenamefont {Clark}, \citenamefont {Schaibley}, \citenamefont {He},
  \citenamefont {Chen}, \citenamefont {Wei}, \citenamefont {Ding},
  \citenamefont {Zhang}, \citenamefont {Yao}, \citenamefont {Xu}, \citenamefont
  {Lu},\ and\ \citenamefont {Pan}}]{nnano.2015.75}%
  \BibitemOpen
  \bibfield  {author} {\bibinfo {author} {\bibfnamefont {Y.-M.}\ \bibnamefont
  {He}}, \bibinfo {author} {\bibfnamefont {G.}~\bibnamefont {Clark}}, \bibinfo
  {author} {\bibfnamefont {J.~R.}\ \bibnamefont {Schaibley}}, \bibinfo {author}
  {\bibfnamefont {Y.}~\bibnamefont {He}}, \bibinfo {author} {\bibfnamefont
  {M.-C.}\ \bibnamefont {Chen}}, \bibinfo {author} {\bibfnamefont {Y.-J.}\
  \bibnamefont {Wei}}, \bibinfo {author} {\bibfnamefont {X.}~\bibnamefont
  {Ding}}, \bibinfo {author} {\bibfnamefont {Q.}~\bibnamefont {Zhang}},
  \bibinfo {author} {\bibfnamefont {W.}~\bibnamefont {Yao}}, \bibinfo {author}
  {\bibfnamefont {X.}~\bibnamefont {Xu}}, \bibinfo {author} {\bibfnamefont
  {C.-Y.}\ \bibnamefont {Lu}},\ and\ \bibinfo {author} {\bibfnamefont {J.-W.}\
  \bibnamefont {Pan}},\ }\bibfield  {title} {\bibinfo {title} {Single quantum
  emitters in monolayer semiconductors},\ }\href
  {https://doi.org/10.1038/nnano.2015.75} {\bibfield  {journal} {\bibinfo
  {journal} {Nat. Nanotechnol.}\ }\textbf {\bibinfo {volume} {10}},\ \bibinfo
  {pages} {497} (\bibinfo {year} {2015})}\BibitemShut {NoStop}%
\bibitem [{\citenamefont {Lohrmann}\ \emph {et~al.}(2015)\citenamefont
  {Lohrmann}, \citenamefont {Iwamoto}, \citenamefont {Bodrog}, \citenamefont
  {Castelletto}, \citenamefont {Ohshima}, \citenamefont {Karle}, \citenamefont
  {Gali}, \citenamefont {Prawer}, \citenamefont {McCallum},\ and\ \citenamefont
  {Johnson}}]{Lohrmann2015}%
  \BibitemOpen
  \bibfield  {author} {\bibinfo {author} {\bibfnamefont {A.}~\bibnamefont
  {Lohrmann}}, \bibinfo {author} {\bibfnamefont {N.}~\bibnamefont {Iwamoto}},
  \bibinfo {author} {\bibfnamefont {Z.}~\bibnamefont {Bodrog}}, \bibinfo
  {author} {\bibfnamefont {S.}~\bibnamefont {Castelletto}}, \bibinfo {author}
  {\bibfnamefont {T.}~\bibnamefont {Ohshima}}, \bibinfo {author} {\bibfnamefont
  {T.}~\bibnamefont {Karle}}, \bibinfo {author} {\bibfnamefont
  {A.}~\bibnamefont {Gali}}, \bibinfo {author} {\bibfnamefont {S.}~\bibnamefont
  {Prawer}}, \bibinfo {author} {\bibfnamefont {J.}~\bibnamefont {McCallum}},\
  and\ \bibinfo {author} {\bibfnamefont {B.}~\bibnamefont {Johnson}},\
  }\bibfield  {title} {\bibinfo {title} {Single-photon emitting diode in
  silicon carbide},\ }\href {https://doi.org/10.1038/ncomms8783} {\bibfield
  {journal} {\bibinfo  {journal} {Nat. Commun.}\ }\textbf {\bibinfo {volume}
  {6}},\ \bibinfo {pages} {7783} (\bibinfo {year} {2015})}\BibitemShut
  {NoStop}%
\bibitem [{\citenamefont {Redjem}\ \emph {et~al.}(2023)\citenamefont {Redjem},
  \citenamefont {Zhiyenbayev}, \citenamefont {Qarony}, \citenamefont {Ivanov},
  \citenamefont {Papapanos}, \citenamefont {Liu}, \citenamefont {Jhuria},
  \citenamefont {Balushi}, \citenamefont {Dhuey}, \citenamefont {Schwartzberg},
  \citenamefont {Tan}, \citenamefont {Schenkel},\ and\ \citenamefont
  {Kanté}}]{Redjem.2023}%
  \BibitemOpen
  \bibfield  {author} {\bibinfo {author} {\bibfnamefont {W.}~\bibnamefont
  {Redjem}}, \bibinfo {author} {\bibfnamefont {Y.}~\bibnamefont {Zhiyenbayev}},
  \bibinfo {author} {\bibfnamefont {W.}~\bibnamefont {Qarony}}, \bibinfo
  {author} {\bibfnamefont {V.}~\bibnamefont {Ivanov}}, \bibinfo {author}
  {\bibfnamefont {C.}~\bibnamefont {Papapanos}}, \bibinfo {author}
  {\bibfnamefont {W.}~\bibnamefont {Liu}}, \bibinfo {author} {\bibfnamefont
  {K.}~\bibnamefont {Jhuria}}, \bibinfo {author} {\bibfnamefont {Z.~Y.~A.}\
  \bibnamefont {Balushi}}, \bibinfo {author} {\bibfnamefont {S.}~\bibnamefont
  {Dhuey}}, \bibinfo {author} {\bibfnamefont {A.}~\bibnamefont {Schwartzberg}},
  \bibinfo {author} {\bibfnamefont {L.~Z.}\ \bibnamefont {Tan}}, \bibinfo
  {author} {\bibfnamefont {T.}~\bibnamefont {Schenkel}},\ and\ \bibinfo
  {author} {\bibfnamefont {B.}~\bibnamefont {Kanté}},\ }\bibfield  {title}
  {\bibinfo {title} {All-silicon quantum light source by embedding an atomic
  emissive center in a nanophotonic cavity},\ }\href
  {https://doi.org/10.1038/s41467-023-38559-6} {\bibfield  {journal} {\bibinfo
  {journal} {Nat. Commun.}\ }\textbf {\bibinfo {volume} {14}},\ \bibinfo
  {pages} {3321} (\bibinfo {year} {2023})}\BibitemShut {NoStop}%
\bibitem [{\citenamefont {Komza}\ \emph {et~al.}(2022)\citenamefont {Komza},
  \citenamefont {Samutpraphoot}, \citenamefont {Odeh}, \citenamefont {Tang},
  \citenamefont {Mathew}, \citenamefont {Chang}, \citenamefont {Song},
  \citenamefont {Kim}, \citenamefont {Xiong}, \citenamefont {Hautier},\ and\
  \citenamefont {Sipahigil}}]{Komza.Sipahigil.2022}%
  \BibitemOpen
  \bibfield  {author} {\bibinfo {author} {\bibfnamefont {L.}~\bibnamefont
  {Komza}}, \bibinfo {author} {\bibfnamefont {P.}~\bibnamefont
  {Samutpraphoot}}, \bibinfo {author} {\bibfnamefont {M.}~\bibnamefont {Odeh}},
  \bibinfo {author} {\bibfnamefont {Y.-L.}\ \bibnamefont {Tang}}, \bibinfo
  {author} {\bibfnamefont {M.}~\bibnamefont {Mathew}}, \bibinfo {author}
  {\bibfnamefont {J.}~\bibnamefont {Chang}}, \bibinfo {author} {\bibfnamefont
  {H.}~\bibnamefont {Song}}, \bibinfo {author} {\bibfnamefont {M.-K.}\
  \bibnamefont {Kim}}, \bibinfo {author} {\bibfnamefont {Y.}~\bibnamefont
  {Xiong}}, \bibinfo {author} {\bibfnamefont {G.}~\bibnamefont {Hautier}},\
  and\ \bibinfo {author} {\bibfnamefont {A.}~\bibnamefont {Sipahigil}},\
  }\bibfield  {title} {\bibinfo {title} {Indistinguishable photons from an
  artificial atom in silicon photonics},\ }\href
  {https://doi.org/https://doi.org/10.48550/arXiv.2211.09305} {\bibfield
  {journal} {\bibinfo  {journal} {arXiv (quant-ph)}\ ,\ \bibinfo {pages}
  {2211.09305v1}} (\bibinfo {year} {2022})},\ \bibinfo {note} {(accessed March
  9, 2022)},\ \Eprint {https://arxiv.org/abs/arXiv:2211.09305}
  {arXiv:2211.09305 [quant-ph]} \BibitemShut {NoStop}%
\bibitem [{\citenamefont {Fournier}\ \emph {et~al.}(2023)\citenamefont
  {Fournier}, \citenamefont {Roux}, \citenamefont {Watanabe}, \citenamefont
  {Taniguchi}, \citenamefont {Buil}, \citenamefont {Barjon}, \citenamefont
  {Hermier},\ and\ \citenamefont {Delteil}}]{Fournier.Delteil.2022}%
  \BibitemOpen
  \bibfield  {author} {\bibinfo {author} {\bibfnamefont {C.}~\bibnamefont
  {Fournier}}, \bibinfo {author} {\bibfnamefont {S.}~\bibnamefont {Roux}},
  \bibinfo {author} {\bibfnamefont {K.}~\bibnamefont {Watanabe}}, \bibinfo
  {author} {\bibfnamefont {T.}~\bibnamefont {Taniguchi}}, \bibinfo {author}
  {\bibfnamefont {S.}~\bibnamefont {Buil}}, \bibinfo {author} {\bibfnamefont
  {J.}~\bibnamefont {Barjon}}, \bibinfo {author} {\bibfnamefont {J.-P.}\
  \bibnamefont {Hermier}},\ and\ \bibinfo {author} {\bibfnamefont
  {A.}~\bibnamefont {Delteil}},\ }\bibfield  {title} {\bibinfo {title}
  {Two-photon interference from a quantum emitter in hexagonal boron nitride},\
  }\href {https://doi.org/10.1103/PhysRevApplied.19.L041003} {\bibfield
  {journal} {\bibinfo  {journal} {Phys. Rev. Appl.}\ }\textbf {\bibinfo
  {volume} {19}},\ \bibinfo {pages} {L041003} (\bibinfo {year}
  {2023})}\BibitemShut {NoStop}%
\bibitem [{\citenamefont {Gao}\ \emph {et~al.}(2023)\citenamefont {Gao},
  \citenamefont {Helversen}, \citenamefont {Antón-Solanas}, \citenamefont
  {Schneider},\ and\ \citenamefont {Heindel}}]{Gao.Heindel.2023}%
  \BibitemOpen
  \bibfield  {author} {\bibinfo {author} {\bibfnamefont {T.}~\bibnamefont
  {Gao}}, \bibinfo {author} {\bibfnamefont {M.~v.}\ \bibnamefont {Helversen}},
  \bibinfo {author} {\bibfnamefont {C.}~\bibnamefont {Antón-Solanas}},
  \bibinfo {author} {\bibfnamefont {C.}~\bibnamefont {Schneider}},\ and\
  \bibinfo {author} {\bibfnamefont {T.}~\bibnamefont {Heindel}},\ }\bibfield
  {title} {\bibinfo {title} {{Atomically-thin single-photon sources for quantum
  communication}},\ }\href {https://doi.org/10.1038/s41699-023-00366-4}
  {\bibfield  {journal} {\bibinfo  {journal} {NPJ 2D Mater. Appl.}\ }\textbf
  {\bibinfo {volume} {7}},\ \bibinfo {pages} {4} (\bibinfo {year}
  {2023})}\BibitemShut {NoStop}%
\bibitem [{\citenamefont {Vogl}\ \emph
  {et~al.}(2019{\natexlab{c}})\citenamefont {Vogl}, \citenamefont {Lecamwasam},
  \citenamefont {Buchler}, \citenamefont {Lu},\ and\ \citenamefont
  {Lam}}]{10.1021/acsphotonics.9b00314}%
  \BibitemOpen
  \bibfield  {author} {\bibinfo {author} {\bibfnamefont {T.}~\bibnamefont
  {Vogl}}, \bibinfo {author} {\bibfnamefont {R.}~\bibnamefont {Lecamwasam}},
  \bibinfo {author} {\bibfnamefont {B.~C.}\ \bibnamefont {Buchler}}, \bibinfo
  {author} {\bibfnamefont {Y.}~\bibnamefont {Lu}},\ and\ \bibinfo {author}
  {\bibfnamefont {P.~K.}\ \bibnamefont {Lam}},\ }\bibfield  {title} {\bibinfo
  {title} {Compact cavity-enhanced single-photon generation with hexagonal
  boron nitride},\ }\href {https://doi.org/10.1021/acsphotonics.9b00314}
  {\bibfield  {journal} {\bibinfo  {journal} {ACS Photonics}\ }\textbf
  {\bibinfo {volume} {6}},\ \bibinfo {pages} {1955} (\bibinfo {year}
  {2019}{\natexlab{c}})}\BibitemShut {NoStop}%
\bibitem [{\citenamefont {Kresse}\ and\ \citenamefont
  {Furthmüller}(1996)}]{vasp1}%
  \BibitemOpen
  \bibfield  {author} {\bibinfo {author} {\bibfnamefont {G.}~\bibnamefont
  {Kresse}}\ and\ \bibinfo {author} {\bibfnamefont {J.}~\bibnamefont
  {Furthmüller}},\ }\bibfield  {title} {\bibinfo {title} {Efficiency of
  ab-initio total energy calculations for metals and semiconductors using a
  plane-wave basis set},\ }\href
  {https://doi.org/https://doi.org/10.1016/0927-0256(96)00008-0} {\bibfield
  {journal} {\bibinfo  {journal} {Comput. Mater. Sci.}\ }\textbf {\bibinfo
  {volume} {6}},\ \bibinfo {pages} {15 } (\bibinfo {year} {1996})}\BibitemShut
  {NoStop}%
\bibitem [{\citenamefont {Kresse}\ and\ \citenamefont
  {Furthm\"uller}(1996)}]{vasp2}%
  \BibitemOpen
  \bibfield  {author} {\bibinfo {author} {\bibfnamefont {G.}~\bibnamefont
  {Kresse}}\ and\ \bibinfo {author} {\bibfnamefont {J.}~\bibnamefont
  {Furthm\"uller}},\ }\bibfield  {title} {\bibinfo {title} {Efficient iterative
  schemes for ab initio total-energy calculations using a plane-wave basis
  set},\ }\href {https://doi.org/https://doi.org/10.1103/PhysRevB.54.11169}
  {\bibfield  {journal} {\bibinfo  {journal} {Phys. Rev. B}\ }\textbf {\bibinfo
  {volume} {54}},\ \bibinfo {pages} {11169} (\bibinfo {year}
  {1996})}\BibitemShut {NoStop}%
\bibitem [{\citenamefont {Bl\"ochl}(1994)}]{paw}%
  \BibitemOpen
  \bibfield  {author} {\bibinfo {author} {\bibfnamefont {P.~E.}\ \bibnamefont
  {Bl\"ochl}},\ }\bibfield  {title} {\bibinfo {title} {Projector augmented-wave
  method},\ }\href {https://doi.org/https://doi.org/10.1103/PhysRevB.50.17953}
  {\bibfield  {journal} {\bibinfo  {journal} {Phys. Rev. B}\ }\textbf {\bibinfo
  {volume} {50}},\ \bibinfo {pages} {17953} (\bibinfo {year}
  {1994})}\BibitemShut {NoStop}%
\bibitem [{\citenamefont {Kresse}\ and\ \citenamefont {Joubert}(1999)}]{paw2}%
  \BibitemOpen
  \bibfield  {author} {\bibinfo {author} {\bibfnamefont {G.}~\bibnamefont
  {Kresse}}\ and\ \bibinfo {author} {\bibfnamefont {D.}~\bibnamefont
  {Joubert}},\ }\bibfield  {title} {\bibinfo {title} {From ultrasoft
  pseudopotentials to the projector augmented-wave method},\ }\href
  {https://doi.org/https://doi.org/10.1103/PhysRevB.59.1758} {\bibfield
  {journal} {\bibinfo  {journal} {Phys. Rev. B}\ }\textbf {\bibinfo {volume}
  {59}},\ \bibinfo {pages} {1758} (\bibinfo {year} {1999})}\BibitemShut
  {NoStop}%
\bibitem [{\citenamefont {Sajid}\ \emph {et~al.}(2020)\citenamefont {Sajid},
  \citenamefont {Reimers}, \citenamefont {Kobayashi},\ and\ \citenamefont
  {Ford}}]{Sajid2020}%
  \BibitemOpen
  \bibfield  {author} {\bibinfo {author} {\bibfnamefont {A.}~\bibnamefont
  {Sajid}}, \bibinfo {author} {\bibfnamefont {J.~R.}\ \bibnamefont {Reimers}},
  \bibinfo {author} {\bibfnamefont {R.}~\bibnamefont {Kobayashi}},\ and\
  \bibinfo {author} {\bibfnamefont {M.~J.}\ \bibnamefont {Ford}},\ }\bibfield
  {title} {\bibinfo {title} {Theoretical spectroscopy of the
  ${\mathrm{v}}_{\mathrm{n}}{\mathrm{n}}_{\mathrm{b}}$ defect in hexagonal
  boron nitride},\ }\href {https://doi.org/10.1103/PhysRevB.102.144104}
  {\bibfield  {journal} {\bibinfo  {journal} {Phys. Rev. B}\ }\textbf {\bibinfo
  {volume} {102}},\ \bibinfo {pages} {144104} (\bibinfo {year}
  {2020})}\BibitemShut {NoStop}%
\bibitem [{\citenamefont {Liu}(2017)}]{PyVaspwfc}%
  \BibitemOpen
  \bibfield  {author} {\bibinfo {author} {\bibfnamefont {L.}~\bibnamefont
  {Liu}},\ }\bibfield  {title} {\bibinfo {title} {Pyvaspwfc.},\ }\href
  {https://github.com/liming-liu/pyvaspwfc} {\bibfield  {journal} {\bibinfo
  {journal} {https://github.com/liming-liu/pyvaspwfc}\ } (\bibinfo {year}
  {2017})},\ \bibinfo {note} {(accessed October 1, 2022)}\BibitemShut {NoStop}%
\end{thebibliography}%



\end{document}